%% file: dynamic.tex
 \documentclass[twocolumn]{aastex631}

\hypersetup{linkcolor=blue,citecolor=blue,filecolor=cyan,urlcolor=blue}

\usepackage{CJK}

\usepackage{longtable}
\usepackage{lipsum} % just for dummy text- not needed for a longtable
\usepackage{multirow}
\usepackage{xspace}

\usepackage{booktabs}

\shorttitle{Dynamics of Embedded Dense Cores}
\shortauthors{Li et al.}
\graphicspath{{./}{figures/}}

\newcommand\ndustcores{294} %415
\newcommand\totalmolcores{129}%164}
\newcommand\cocores{267} %262}
\newcommand{\HII}{H{\scriptsize II}\xspace}
\newcommand{\um}{{$\mu$m}\xspace}
\def\arcsec{$^{\prime\prime}$\xspace}
\def\Mo{$M_{\odot}$\xspace}
\def\kms{km~s$^{-1}$\xspace}
\def\co18{C$^{18}$O}
\def\dcop{DCO$^{+}$\xspace}
\def\n2dp{N$_{2}$D$^{+}$}

\def\ch3oh{CH$_{3}$OH}
\def\h2co{H$_{2}$CO}
\def\1{\uppercase\expandafter{\romannumeral1}}
\def\2{\uppercase\expandafter{\romannumeral2}}
\def\3{\uppercase\expandafter{\romannumeral3}}
\def\6{\uppercase\expandafter{\romannumeral6}}
\def\7{\uppercase\expandafter{\romannumeral7}}
\def\8{\uppercase\expandafter{\romannumeral8}}
\def\9{\uppercase\expandafter{\romannumeral9}}

\begin{document}

\title{The ALMA Survey of 70 $\mu$m Dark High-mass Clumps in Early Stages (ASHES). \8. Dynamics of Embedded Dense Cores}

\correspondingauthor{Shanghuo Li}
\email{shanghuo.li@gmail.com,li@mpia.de}

\input{author}
% \author[0000-0003-1275-5251]{Shanghuo Li }
%\affiliation{Max Planck Institute for Astronomy, K\"onigstuhl 17, 69117 Heidelberg, Germany}

\begin{abstract}
We present the dynamical properties of 294 cores 
embedded in twelve IRDCs  
observed as part of the ASHES Survey.  
Protostellar cores have higher gas masses, 
surface densities, column densities, and volume 
densities than prestellar cores, indicating 
core mass growth from the prestellar to the protostellar 
phase. We find that $\sim$80\% of cores with virial 
parameter ($\alpha$) measurements are gravitationally 
bound ($\alpha<$ 2).  We also find an anticorrelation 
between the mass and the virial parameter of cores, 
with massive cores having on average lower virial 
parameters. Protostellar cores are more gravitationally 
bound than prestellar cores, with an average virial 
parameter of 1.2 and 1.5, respectively. 
The observed nonthermal velocity dispersion 
(from  N$_{2}$D$^{+}$ or DCO$^{+}$)  
is consistent with simulations in which 
turbulence is continuously injected, whereas the 
core-to-core velocity dispersion is 
neither in agreement with driven nor decaying turbulence 
simulations. We find a no significant increment in the line 
velocity dispersion from prestellar to protostellar cores, 
suggesting that the dense gas within the core traced 
by these deuterated molecules is not yet severely affected by 
turbulence injected from outflow activity at the early 
evolutionary stages traced in ASHES. The most massive 
cores are strongly self-gravitating and have 
greater surface density, Mach number, and velocity 
dispersion than cores with lower masses. 
Dense cores do not have significant velocity shifts 
relative to their low-density envelopes, suggesting that 
 dense cores are comoving with their envelopes. 
 We conclude that the observed core properties are 
 more in line with the predictions of ``clump-fed" scenarios 
 rather than those of ``core-fed" scenarios.

\end{abstract}


\section{Introduction} 
\label{sec:intro}
High-mass stars ($M\gtrsim8\,M_\odot$) are mainly responsible 
for the chemical enrichment and kinetic energy injection 
into the interstellar matter (ISM) of galaxies \citep{Kennicutt-1998,McKee-2007,
Zinnecker-2007}.  
A great progress has been made in investigating the basic 
properties of high-mass star-forming regions. 
However, the formation of high-mass stars is not yet well 
understood, in particular their earliest stages of evolution. 
High-mass stars are known to form from dense 
($\sim$$10^6$ cm$^{-3}$), compact ($\sim$$0.01-0.1$ pc) 
self-gravitating regions, or `cores', that are embedded in 
more extended molecular clumps ($\lesssim$1 pc)
which are dense substructures residing in molecular clouds 
\citep[10--100 pc;][]{Bergin-2007,Zhang-2009,Zhang-2015}. 
A gravitationally bound dense core prior to the protostellar 
phase is called a prestellar core. A prestellar core  
has not yet formed any central protostellar object and 
is on the verge of gravitational collapse 
\citep{1994MNRAS.268..276W,Redaelli-2021}.

Prestellar cores represent the starting point in the star 
formation process. Different high-mass star 
formation theories predict very different initial conditions 
for the prestellar cores.  In the ``core-fed" or 
``core-accretion" models, a molecular cloud 
will fragment into cores of different masses. The most 
massive cores will collapse and form high-mass stars, 
while the low-mass cores will evolve into low-mass stars 
\citep[e.g.,][]{mckee-tan-2002}. In these models, the 
high-mass prestellar cores are formed quasi-statically. 
On the other hand, in the ``clump-fed" scenarios, 
prestellar cores initially 
have masses comparable to the Jeans mass,  are
 subvirialized, and will accrete material from the reservoir 
 of gas of the parental cloud \citep[e.g.,][]{Bonnell2001,bonnell-2004,
2009MNRAS.400.1775S,2019MNRAS.490.3061V,Padoan-2020,Pelkonen-2021}.

For these models, the turbulent feedback in molecular 
clouds also affects start formation. \citet{Offner-2008},  \citet{2005Natur.438..332K} suggest that the presence 
or absence of turbulent feedback is directly related to 
the star formation mechanism. If the turbulence is 
maintained in the cloud, then the mass of the cores is 
limited by the initial turbulent compression. 
On the other hand, if turbulence decays quickly, then the 
virial parameter decreases significantly, and competitive 
accretion might be possible in the clouds. 

Since high-mass stars evolve very quickly, compared to 
their low-mass counterparts, and rapidly change their 
environment, in order to determine their formation 
mechanisms, it is necessary to study high-mass 
star-forming regions in a very early stage of evolution. 
The perfect test beds to study different star formation 
mechanisms are infrared dark clouds (IRDCs). 
IRDCs are molecular clouds that appear dark at 
mid-infrared wavelengths against the strong Galactic 
background emission \citep{egan-1998,simon-2006}. 
They have high column densities 
($\sim$10$^{22}$ cm$^{-2}$), and are considered to be 
the birth places of high-mass stars \citep{Rathborne-2004}.

Recently, there have been a handful of observations 
aimed to determine the properties of the gas and 
dust toward IRDCs \citep[e.g.,][]{Bontemps-2010,
Sanhueza-2010,Sanhueza-2012,Sanhueza-2013,Palau-2013,
Wang2014,Sanhueza-2017,Lu2015,Contreras-2016,Csengeri-2017, 
Henshaw-2017, Cyganowski-2017,Pillai-2019,
Li-2019b,Li-2020a,Li-2021,Barnes2021}. 
However, most of them lack sufficient angular resolution 
to identify individual star-forming cores, or they target 
sources with embedded infrared emission, suggesting 
that active star formation might have already started.

In this paper, we present sensitive, high-angular 
resolution ($\sim$0.023 pc) and high-sensitivity 
observations of the gas and dust emission within 
twelve 70 $\mu$m dark IRDCs from the Atacama Large Millimeter/submillimeter (ALMA) Survey 
of 70 $\mu$m dark High-mass clumps in Early 
Stages \citep[ASHES;][]{Sanhueza-2019}. 
We study the gas kinematics 
of dense embedded cores revealed by the 
continuum emission in \cite{Sanhueza-2019}.  
These IRDCs were selected from the Millimetre 
Astronomy Legacy Team survey 
\citep[MALT90;][]{jackson-2013,Foster-2013,
Foster-2011}, which targeted sources from the 
Atacama Pathfinder Experiment (APEX) Telescope 
Large Area Survey of the GALaxy 
\citep[ATLASGAL;][]{schuller-2009,Contreras-2013a}. 
They are located at distances within 5.5 kpc 
\citep{Whitaker-2017}, and have the perfect condition 
of cluster-forming clumps in early stages of evolution: 
they have large masses   
\citep[$>500\,M_\odot$;][]{Contreras-2017}, large 
surface densities ($>0.1$ g~cm$^{-2}$), and large 
volume densities ($>10^4$ cm$^{-3}$) ensuring that 
they will form high-mass stars. They also have low 
dust temperatures at the clump scale 
\citep[$\leq 15$ K;][]{Guzman-2015}, and they appear 
dark from 3.6 to 70 $\mu$m in \textit{Spitzer/Herschel}, 
suggesting that they have no yet embedded powerful internal 
heating sources. More details on the sample 
selection can be found in \citet{Sanhueza-2019}. 
In spite of being IR dark even up to 70 $\mu$m, most 
of these twelve IRDCs have deeply embedded star 
formation activity, as revealed by ``warm core" tracers
\citet{Sanhueza-2019} and molecular 
outflows \citep{Li-2020a,Tafoya-2021,Morii-2021}. 
For statistical details of the outflow content in these 
IRDCs, the reader can refer to \citet{Li-2020a}. 
The first work on the detailed deuterated chemistry, 
as a first step on a single target, is presented in 
\cite{Sakai-2021}.  The chemistry and the CO 
depletion fraction of all the detected ASHES cores 
from the pilot survey \citep{Sanhueza-2019} 
are presented in \cite{Li-2022b} and 
\cite{Sabatini-2022}, respectively. 
The whole ASHES survey of thirty-nine 70 $\mu$m 
dark clumps is presented in a companion 
paper \citep{Morii-2023}.

In this work, using the molecular line and continuum 
emission, we study the properties of embedded dense 
cores. The observations are described in 
Section~\ref{sec:obs}. In Section~\ref{sec:res} we 
present our results and analysis. We discuss the 
kinematical properties of embedded dense cores in 
Section~\ref{sec:dis}.  Finally, Section~\ref{sec:con} 
presents the summary of our findings.

\begin{deluxetable*}{llccccccccccccccc}
\caption{Summary of Clumps Properties.}\label{summaryclumps}
\tabletypesize{\scriptsize}
\tablehead{
\colhead{Name} & \colhead{Abbreviation}& \colhead{Dist.} & \colhead{$M_{\rm gas}$} & $M_{\rm vir}$ & \colhead{$\alpha$} & \colhead{$v_{\rm LSR}$} & \colhead{$\sigma_{\rm obs}$}  \\
 \colhead{}&\colhead{}& \colhead{(kpc)} & \colhead{10$^3$ ($M_\odot$)} & \colhead{10$^3$ ($M_\odot$)} &\colhead{} & \colhead{(km s$^{-1}$)}& \colhead{(km s$^{-1}$)}}
\startdata
G010.991-00.082 & G10.99 & 3.7 & 2.2$\pm$0.6 & 1.1$\pm$0.2 & 0.5$\pm$0.2 & 29.53 & 1.27$\pm$0.05  \\
G014.492-00.139 & G14.49 & 3.9 & 5.2$\pm$1.1 & 3.0$\pm$0.4 & 0.6$\pm$0.1 & 41.14 & 1.68$\pm$0.05 \\
G028.273-00.167 & G28.27 & 5.1 & 1.5$\pm$0.7 & 0.4$\pm$0.1 & 0.3$\pm$0.1 &80.00 &1.34$\pm$0.13 \\
G327.116-00.294 & G327.11 & 3.9 & 0.6$\pm$0.1 & 0.1$\pm$0.1 & 0.2$\pm$0.1 & -58.93 & 0.56$\pm$0.05  \\
G331.372-00.116 & G331.37 & 5.4 & 1.6$\pm$0.2 & 1.3$\pm$0.2 & 0.8$\pm$0.2 & -87.83 & 1.29$\pm$0.05 \\
G332.969-00.029 & G332.96 & 4.4 & 0.7$\pm$0.1 & 0.7$\pm$0.2 & 1.0$\pm$0.3 & -66.63 & 1.41$\pm$0.05  \\
G337.541-00.082 & G337.54 & 4.0 & 1.2$\pm$0.2 & 2.8$\pm$0.7 & 2.4$\pm$0.7 & -54.57 &2.00$\pm$0.05 \\
G340.179-00.242 & G340.17 & 4.1 & 1.5$\pm$0.3 & 1.8$\pm$0.2 & 1.2$\pm$0.3 & -53.69 &1.48$\pm$0.05 \\
G340.222-00.167 & G340.22 & 4.0 & 0.8$\pm$0.1 & 4.4$\pm$1.0 & 5.7$\pm$1.7 & -51.30 &3.04$\pm$0.05 \\
G340.232-00.146 & G340.23 & 3.9 & 0.7$\pm$0.1 & 0.7$\pm$0.2 & 1.0$\pm$0.3 & -50.78 &1.23$\pm$0.05 \\
G341.039-00.114 & G341.03 & 3.6 & 1.1$\pm$0.2 & 0.7$\pm$0.1 & 0.6$\pm$0.2 & -43.04 &0.97$\pm$0.05 \\
G343.489-00.416 & G343.48 & 2.9 & 0.8$\pm$0.3 & 0.4$\pm$0.1 & 0.5$\pm$0.2 & -28.96 &1.00$\pm$0.05 \\
\enddata
\tablenotetext{}{N$_{2}$H$^{+}\, J$ = 1--0 emission was used to derive 
the velocity dispersion for 
the clumps, except for G028.273-00.167,  
G337.541--00.082, and G340.222--00.167. 
The former clump was obtained using 
NH$_2$D J$_{K_a,K_c} = 1_{1,1}\rightarrow1_{0,1}$ 
emission, and the latter two clumps were derived from  
HNC $J$ = 1--0 emission. }
\end{deluxetable*}

\section{Observations}
\label{sec:obs}
The observations were carried out with the Atacama 
Large Millimeter/submillimter  Array (ALMA), located 
in the Llano de Chajnantor, Chile, during the ALMA 
cycle (3) and (4) (project ID: 2015.1.01539.S, 
PI: Sanhueza). These observations covered both the 
continuum and molecular line emission toward 12 
massive 70 $\mu$m dark  high-mass clumps located 
in the Galactic plane. 
Table \ref{summaryclumps} shows the values  
for the masses ($M_{\rm gas}$), virial masses 
($M_{\rm vir}$), virial parameter ($\alpha$), 
source velocity ($v_{\rm LSR}$), and velocity 
dispersion ($\sigma_{\rm obs}$) of the molecular 
line 
emission for the 12 IRDC clumps obtained from 
single-dish telescopes  \citep{Contreras-2017,
Rathborne-2016}, and indicates whether there is 
any asymmetries in the line profiles. 

The 12 m array observations had between 36 
and 48 antennas, with baselines ranging from 
15 to 704 m. Large-scale dust continuum and 
line emission was also recovered thanks to the 
inclusion of the 7 m array and the total power 
(TP; only for line emission) antennas. 
The 7 m array observations consisted of 8--10 
antennas, with baselines ranging from 8 to 44 m. 
Each clump was covered by a 10 point mosaic 
using the 12 m array and by a 3 pointing mosaic 
using the 7 m array, except for G028.27 that was 
observed with 11 and 5 pointing, respectively.  
The angular resolution of the images is $\sim$1\farcs2, 
corresponding to 4800 au (or 0.023 pc) at the 
averaged source distance of 4 kpc.  
The primary beams at 224 GHz of the 12  and 7 m 
arrays are 25\farcs2 (0.49 pc at the 
averaged source distance of 4 kpc) and 44\farcs6 
(0.86 pc at the averaged source distance of 4 kpc), 
respectively. 
The maximum scale recovered scale of the 7 m array 
at the observed frequency is about 30\arcsec 
(0.58 pc at the averaged source distance of 4 kpc).  

The receiver was set to the band 6 of ALMA, 
centered at $\sim$224 GHz in dual polarization mode. 
The velocity resolution of the spectral windows 
ranged between 0.17 and 
1.3 \kms. 
In this paper, we mostly focus on the \n2dp (3-2), 
\dcop (3-2), and C$^{18}$O (2-1) molecular transitions,  
whereas the \h2co and \ch3oh are also used to study 
the core-to-core velocity dispersion. 
The former two lines (\n2dp and \dcop) and 
latter three lines (C$^{18}$O, \h2co, and \ch3oh) 
have spectral resolutions of 0.17 and 
1.3 \kms, respectively. 
Calibration of the observations was carried out 
using the Common Astronomy Software 
Applications (CASA) software package version 
4.5.3, 4.6, and 4.7, while imaging was done 
using CASA 5.4 \citep{McMullin-2007}.

The 12  and 7 m array datasets were 
concatenated and imaged together using the 
CASA 5.4 \texttt{tclean} algorithm. Line cubes 
were made using the \texttt{yclean} script, which 
automatically clean each map channel with 
custom-made masks \citep{Contreras-2018}. 
Natural weighting was used. To avoid artifacts 
due to the complex structure of the IRDCs 
emission, we used a multiscale clean, with scale 
values of 1, 3, 10, and 30 times the image pixel size 
of 0\farcs2. The 12 and 7 m array line emission was 
combined with the TP observations through the 
feathering technique. The achieved line rms is 
$\sim$0.06 K ($\sim$3.6 mJy beam$^{-1}$) and $\sim$0.02 K 
($\sim$1.2 mJy beam$^{-1}$) at a velocity resolution 
of 0.17 \kms and 1.3 \kms, respectively. 
More details on the observations can be found in 
\cite{Sanhueza-2019}.

\section{Results and Analysis}
\label{sec:res}

\subsection{Core Sample}
\label{sec:cores}
For each clump, \citet{Sanhueza-2019} determined 
their embedded cores using the astropy 
\texttt{dendrogram} package 
\citep{Rosolowsky-2008,astropy:2018}. 
A dendrogram was applied 
to both the emission from the image obtained using 
the data of the 12~m array alone, and to the images 
obtained from the 12 and 7~m arrays combined. 
\citet{Sanhueza-2019} determined that by including 
the 7 m array data it was possible to detect fainter 
cores within the IRDCs, thus allowing to trace a 
wider range of core masses. Therefore, for this 
paper, we use the cores detected in the images 
obtained by combining the 12 and 7 m array datasets.
In total, there are \ndustcores~cores in the 12 clumps.

\subsection{Core Evolutionary Stage Classification}
\label{sec:calss}
We have used the classification of  cores into prestellar core 
candidates (hereafter prestellar core) and protostellar 
cores defined in \cite{Sanhueza-2019}, but further 
refined in \cite{Li-2022b}. 
This classification is based on whether there is 
a detection of warm gas tracers, 
i.e., CH$_{3}$OH $4_{2,2}-3_{1,2}$ 
($E_{u}/k$ = 45.46 K; where $k$ is 
Boltzmann's constant), 
H$_{2}$CO $3_{2,2}-2_{2,1}$ ($E_{u}/k$ = 68.09 K), 
and H$_{2}$CO $3_{2,1}-2_{2,0}$ 
($E_{u}/k$ = 68.11 K),  
or if the cores have outflows detected in the CO, 
SiO, and/or H$_2$CO molecular line emission. 
A core is classified as prestellar core if it is 
not associated with emission from any of three 
aforementioned lines nor molecular outflow 
signatures. On the contrary, cores presenting any of 
three aforementioned lines and/or molecular outflows 
are classified as protostellar.  In total,  
we classified the core sample into 97 protostellar 
cores and 197 prestellar cores (category (1)). 
The protostellar cores can further classify into three 
subcategorizes, 
1) ``outflow core" (category (2)) if it is associated with outflows but 
without detection of any of three aforementioned warm 
lines, 2) ``warm core" (category (3)) if it is associated with any of three 
aforementioned warm lines but without outflow detection, 
and 3) ``warm \& outflow core" (category 
(4)) if it is associated with both 
outflows and any of three aforementioned warm lines 
\citep[see also][]{Sanhueza-2019,Li-2022b}.

We note that, although some cores are classified 
as protostellar, they still have no emission detected at 
infrared wavelengths; thus, they are in an earlier 
evolutionary stage than the cores traditionally 
classified as protostellar. 
This classification scheme will be used throughout 
the paper to determine whether any of the properties 
derived for the cores change with their evolutionary 
stage.

\addtolength{\tabcolsep}{-1.95pt}    
\begin{deluxetable*}{ccccccccccccc}
\tabletypesize{\scriptsize} %%\scriptsize \ tiny
\tablecolumns{13} 
\tablewidth{0pt}
\tablecaption{Summary of Dense Cores properties
\label{table_cores}}
\tablehead{ 
\colhead{Clump} 		& \colhead{Core} 			&  
\colhead{$M_{\rm gas}$}	& \colhead{$R$}  				& \colhead{$N_{\rm peak}$(H$_{2}$)} 	&  \colhead{$\Sigma$} 		& \colhead{T$_{\rm NH_3}$} 	& \colhead{$\sigma_{\rm tot}$}	& \colhead{$M_{\rm vir}$}	 	& 
\colhead{$\alpha$} 		& \colhead{$a_{\rm G}$} 		& \colhead{$a_{\rm K}$} 		&  \colhead{SF}   \\
					& 						&  
\colhead{($M_{\rm \odot}$)} & \colhead{(pc)}   			& \colhead{($\times 10^{23}$cm$^{-2}$)} &  
\colhead{(g cm$^{-2}$)}	& \colhead{(K)} 				& \colhead{(km s$^{-1}$)} 		& \colhead{($M_{\rm \odot}$)} 	& 
					& \colhead{(pc Myr$^{-2}$)}	&  \colhead{(pc Myr$^{-2}$)} 	& 
} 
\decimalcolnumbers
\startdata
G10.99 & 1 & 6.90  & 0.024  & 1.62  & 0.78  & 13.4  & 0.34  & 2.47  & 0.36  & 10.54  & 5.09  & 3	\\
G10.99 & 2 & 1.68  & 0.009  & 1.53  & 1.41  & 12.5  & 0.34  & 0.78  & 0.46  & 18.99  & 13.45  & 1	\\
G10.99 & 3 & 3.10  & 0.013  & 1.57  & 1.27  & 12.1  & 0.47  & 2.14  & 0.69  & 17.19  & 18.49  & 1	\\
G10.99 & 4 & 2.05  & 0.014  & 1.06  & 0.70  & 14.2  & 0.24  & 0.66  & 0.32  & 9.52  & 4.48  & 3	\\
G10.99 & 5 & 3.24  & 0.015  & 1.16  & 0.96  & 10.9  & 0.25  & 0.66  & 0.20  & 13.04  & 4.44  & 0	\\
G10.99 & 6 & 0.60  & 0.007  & 0.82  & 0.78  & 13.2  & ... & ... & ... & 10.57  & ... & 3	\\
G10.99 & 7 & 2.87  & 0.016  & 0.75  & 0.72  & 12.2  & 0.27  & 0.98  & 0.34  & 9.69  & 4.58  & 0	\\
G10.99 & 8 & 2.17  & 0.017  & 0.68  & 0.52  & 12.3  & 0.35  & 1.37  & 0.63  & 7.08  & 7.70  & 0	\\
G10.99 & 9 & 1.44  & 0.012  & 0.66  & 0.71  & 11.5  & 0.40  & 1.34  & 0.93  & 9.58  & 14.04  & 0	\\
G10.99 & 10 & 0.92  & 0.013  & 0.55  & 0.35  & 12.8  & 0.34  & 1.04  & 1.12  & 4.79  & 8.97  & 1	\\
G10.99 & 11 & 0.53  & 0.007  & 0.62  & 0.79  & 11.8  & 0.22  & 0.22  & 0.41  & 10.65  & 7.59  & 1	\\
\enddata
(This table is available in its entirety in a machine-readable form in the online
journal. A portion is shown here for guidance regarding its form and content.)
\tablenotetext{}{Summary of the molecular line emission and virial properties derived for all the cores embedded in the 12 IRDCs observed with ALMA. Columns (1) and (2) show the short name of the parental IRDC and dense cores, respectively. 
The gas mass, core radius, peak H$_{2}$ density, core-averaged surface density are presented in columns (3)--(6). Column (7) shows the gas temperature derived from NH$_{3}$. 
The total velocity dispersions derived from \n2dp or 
\dcop, virial mass, and virial parameters are presented 
in columns (8)--(10), respectively. 
Columns (11) and (12) are the $a_{\rm G}$ and $a_{\rm K}$, 
respectively.  
Columns (13) shows the star formation category defined for each core, where ``0" means no star formation signature (prestellar core candidate), ``1" represents the outflow core, ``2" is warm core, and ``3" means warm and outflow core.  
}
\end{deluxetable*}

%----------------------------------------------------
\begin{figure*}[!ht]
\centering
\includegraphics[scale=0.25]{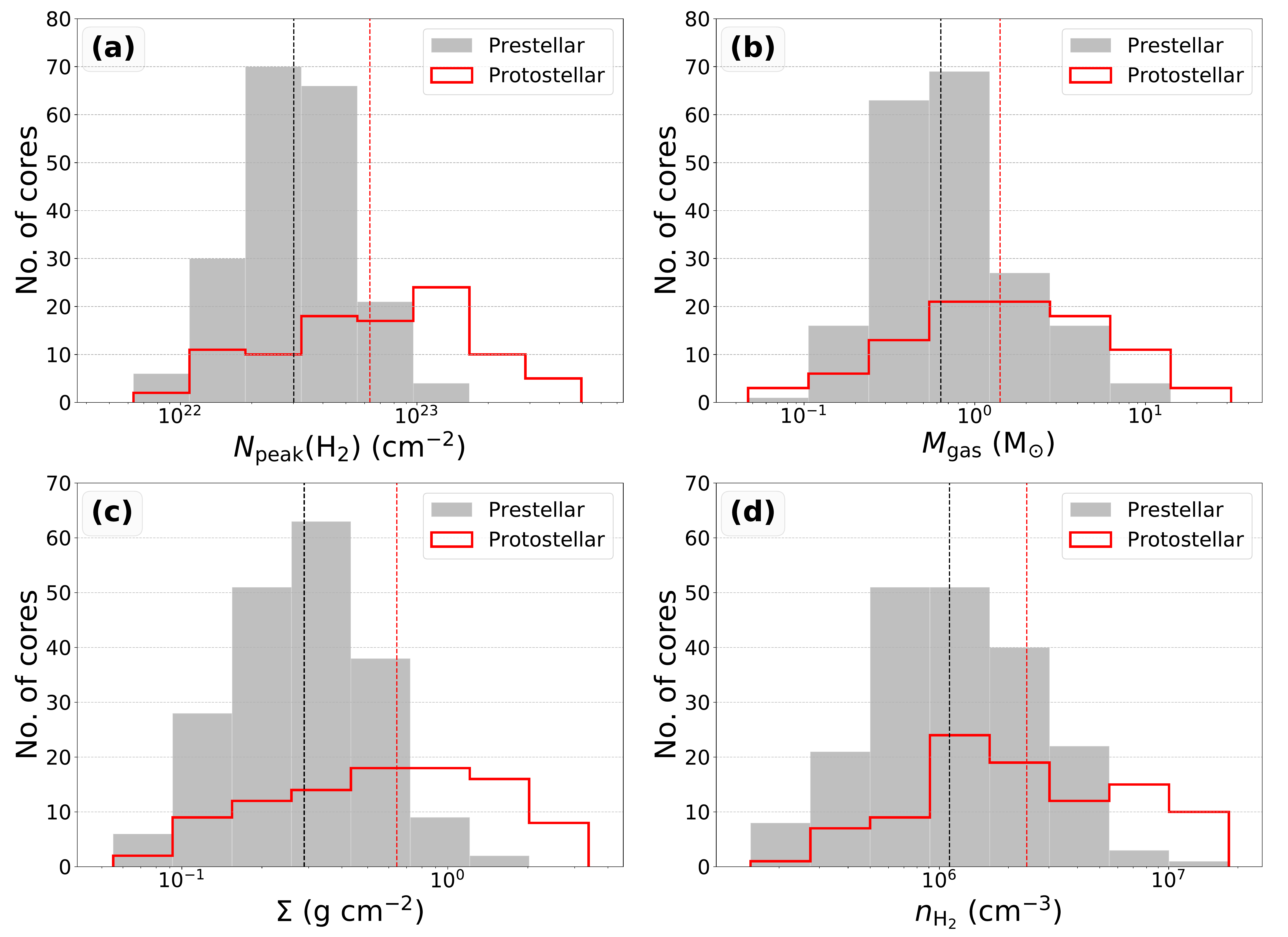}
\caption{
Panels (a)--(d) show the histograms of the peak column 
density ($N_{\rm peak}$(H$_{2}$)), 
gas mass ($M_{\rm gas}$), 
surface density ($\Sigma_{\rm H_2}$), 
and volume density ($n_{\rm H_2}$)  
for the prestellar and protostellar cores. 
The black and red dashed vertical lines indicate the median 
values of prestellar and protostellar cores, respectively. 
}
\label{fig:NH2}
\end{figure*}
%%----------------------------------------------------

\subsection{Core Mass and Core Density}
\label{sec:mass}
We retrieved the rotational excitation temperature 
($T_{\rm NH_3}$), that is derived from NH$_{3}$ (1,1) 
and (2,2) transition lines \citep[see Appendix B in][for 
detailed procedure on excitation temperature 
determination]{Li-2022b},  from the Complete 
ATCA\footnote{The Australia Telescope Compact Array} 
Census of High-Mass Clumps (CACHMC; Allingham et al., 2023, in preparation) 
at about 5\arcsec angular resolution.  
The retrieved $T_{\rm NH_3}$ is used as excitation 
temperature in the calculation of all molecular parameters 
and core mass,  except for G332.96 that has no available 
NH$_{3}$ data. 
The dust temperature of 12.6~K derived from the clump 
scale is used for G332.96 \citep{Sanhueza-2019}.  
The prestellar and protostellar cores have similar  
$T_{\rm NH_3}$, with the median values of 14.1  
and 14.8 K for the prestellar and protostellar 
cores, respectively. 
These values are lower than the typical gas 
temperature $\geqslant$ 20 K  of protostellar cores 
reported in previous studies 
\citep[e.g.,][]{2014ApJ...790...84L,
2019MNRAS.483.3146B}. 
This further supports  the notion that these 
protostellar cores are still at a very early evolutionary 
phase. 
The low temperature in the protostellar cores could 
be attributed to the fact that these cores are still at a very 
early evolutionary stage, and the surrounding 
core materials have not been sufficiently warmed up. 
On other other hand, we cannot fully rule out the 
possibility that the NH$_{3}$ observations have not  
enough sensitivity or angular resolution to reveal 
local small temperature enhancement toward the 
protostellar cores. 
%have some observational bias regarding the $T_{\rm NH_3}$
%angular resolution is coarse than our ALMA  
%observations ($\sim$1\farcs2)

We have updated the gas mass ($M_{\rm gas}$), 
peak column density ($N_{\rm peak}$(H$_{2}$)), 
volume density ($n_{\rm H_2}$), 
and surface density 
($\Sigma = M_{\rm gas}/(\pi r^{2})$) using the 
temperatures derived from the NH$_{3}$ data 
\citep[see][]{Li-2022b}.   The updated 
parameters are 0.05--31.66 \Mo for the gas mass,  
6.3\,$\times$\,10$^{21}$--5.0\,$\times$\,10$^{23}$ cm$^{-2}$ 
for the peak column density,
1.5\,$\times$\,10$^{5}$--1.8\,$\times$\,10$^{7}$ cm$^{-3}$ 
for the core-averaged volume density, and 
0.05--3.39 g cm$^{-2}$ for the core-averaged 
surface density.  
The updated parameters are slightly  different compared 
to the previous results using the clump-scale dust 
temperatures in \cite{Sanhueza-2019}. The updated 
parameters of each core are tabulated in 
Table~\ref{table_cores} (see also Figure~\ref{fig:NH2}).

\subsection{Properties of the Molecular Line Data}
%\label{sec:}
%
Due to the low temperatures ($<$20 K) and high 
densities ($>10^5$ cm$^{-3}$) found in cores at  
early stages of evolution, there is a high level of 
deuteration due to freeze out of the CO molecules 
into dust grains \citep[e.g.,][]{Caselli-2002,Tan-2017,Sabatini-2019,Sabatini-2022,Li-2021}. 
This makes the emission from the \n2dp ($J$=3-2;  
$n_{\rm crit}=1.7\times10^6$ cm$^{-3}$) and 
\dcop ($J$=3-2; $n_{\rm crit}=1.8\times10^6$ cm$^{-3}$) 
molecules almost exclusively associated to the cold and 
dense gas toward the cores. Unlike \h2co and \ch3oh 
that are associated with more turbulent gas components, 
\n2dp and \dcop preferentially trace quiescent dense 
gas \citep[see][]{Li-2022b}. 
Therefore, \n2dp and \dcop are excellent tracers of the 
gas velocity dispersion in cold dense cores, minimizing 
the contribution from more diffuse intra-clump gas and/or 
more turbulent gas related to protostellar activity, like, for example, molecular outflows.

To determine the properties of the core envelope, defined 
as the lower-density gas surrounding a dense core, we 
used the C$^{18}$O emission. C$^{18}$O ($J$=2-1; 
$n_{\rm crit}=9.3\times10^{3}$ cm$^{-3}$) traces 
lower-density gas;   
thus by determining the properties of the emission 
observed in the same line of sight to the cores, we can 
have an idea of the material that might be associated 
to the gas surrounding each core, or core envelope.

For each core detected in the continuum, we extracted 
the core-averaged spectrum of the \n2dp, \dcop,  
C$^{18}$O, \h2co, and \ch3oh molecular line emission 
within the same area defined for each core by the 
dendrogram technique in \cite{Sanhueza-2019}. 
The spectral extraction was done for the images 
created by combining the 12 and 7 m arrays 
(hereafter 12m7m or interferometric images), 
and for the images created by 
adding the TP array data to the 12m7m dataset 
via the feathering technique (hereafter 12m7mTP or 
feathered images). We preformed Gaussian fittings 
to the core-averaged spectrum for each line, 
except for \n2dp that is fitted using hyperfine line 
structures (hfs); 
the detail fitting processes can be found in \cite{Li-2022b}. 
The best-fit parameters, including the peak brightness 
temperature ($I$), the line central velocity 
($v_{\rm LSR}$), and observed velocity dispersion 
($\sigma_{\rm obs}$ = FWHM$_{\rm obs}$/$\rm 2\sqrt{2ln2}$),  
are presented in Table~\ref{tabla_lines}.

The C$^{18}$O emission, along the line of sight, 
presents asymmetric profiles toward some of cores. 
These line features in our sample can 
be attributed mostly to multiple velocity components 
because the C$^{18}$O is virtually always optically thin 
in this sample \citep[see][]{Sabatini-2022}, and these  
multiple velocity components match those 
of N$_{2}$H$^{+}$ ($J$=1-0) and 
ortho-H$_{2}$D$^{+}$ (1$_{1,0}$-1$_{1,1}$) 
toward G14.49 
\citep[the latter two lines are from][]{Redaelli2022}.  
In order to fit the C$^{18}$O emission associated to 
the cores, the central velocity derived 
from the other dense gas tracers 
(i.e., \dcop, \n2dp, \ch3oh, or \h2co) 
was used to guide the Gaussian fit of the 
C$^{18}$O line.  
 
%Therefore, the C$^{18}$O 
%line parameters obtained via Gaussian fittings might be 
%sometimes under- or over-estimated.  

\begin{figure*}
\centering
\includegraphics[clip,width=0.95\textwidth]{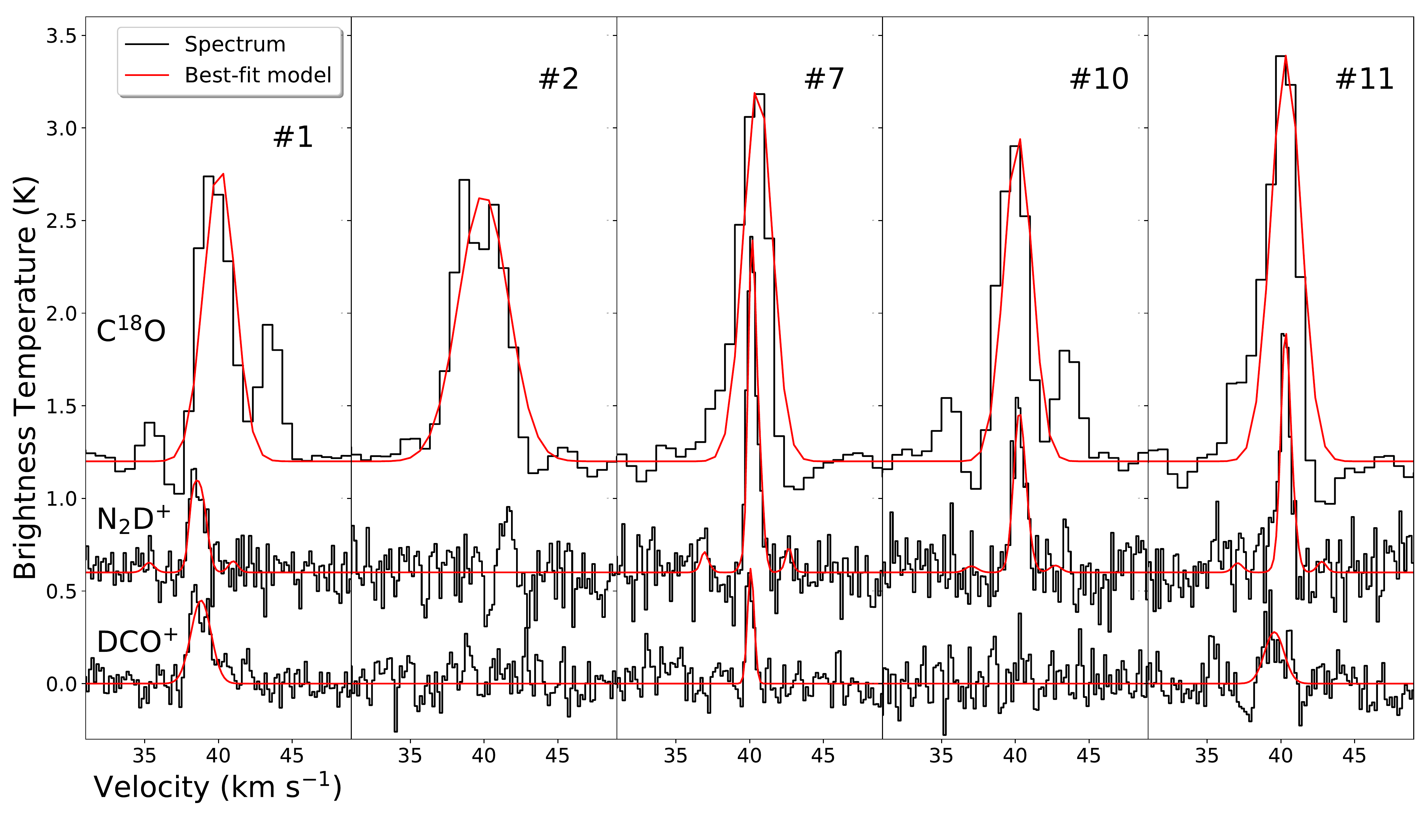}
\caption{Examples of spectra of C$^{18}$O, 
\n2dp, and \dcop for five selected cores in 
the G14.49. 
The results of best fit are overlaid on the 
spectra as red curves. 
}
\label{examplefit}%
\end{figure*}

Table \ref{tabla_lines} summarizes the properties derived from the Gaussian fit for 
all cores, and Figure \ref{examplefit} shows examples  
of the molecular lines emission and the fits for five 
selected cores.

%%---------------------------------------------------
%%---------------------------------------------------

%%---------------------------------------------------
\begin{figure*}
\centering
\includegraphics[trim={0.cm 0cm 0.cm 0cm}, 
clip,width=0.85\textwidth]{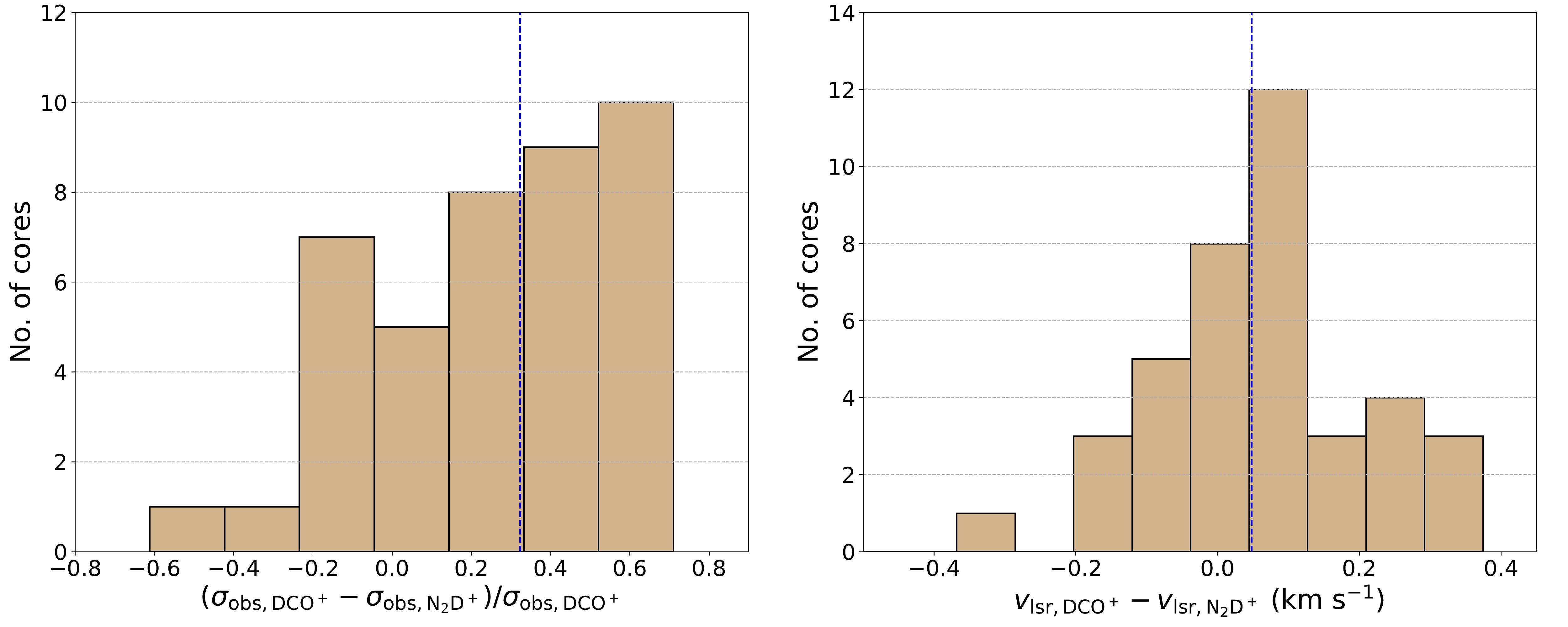}
\caption{Left panel: histogram of the differences 
between the velocity dispersion of the  \dcop and 
\n2dp emission. 
Right panel: histogram of the differences between 
the $v_{LSR}$ of these two molecules. In both 
panels, the histogram shows the differences based 
on the spectra from the 12  and 7 m images.
The blue dashed line indicates the median value 
in each panel. 
}
\label{fig:diff}%
\end{figure*}
%%---------------------------------------------------

%%%----------------------------------------------------
\addtolength{\tabcolsep}{1.8pt}    
\begin{deluxetable*}{cccccccccccc}
\tabletypesize{\scriptsize} %% \ tiny
\tablecolumns{21} 
\tablewidth{0pt}
\tablecaption{Summary of \n2dp and \dcop lines properties
\label{tabla_lines}}
\tablehead{ 
\colhead{Clump} 			& \colhead{Core} 			& 					&   					 & 
\n2dp					&  					 & 				&  						& 							& \dcop				 	& 				 & 				  \\
\cmidrule(r){3-7} \cmidrule{8-12}
						&  						&\colhead{$T_a$} 		&\colhead{$v_{\rm LSR}$}		& 
\colhead{$\sigma_{\rm obs}$} 	& \colhead{$\sigma_{\rm nth}$} 	&\colhead{$\mathcal{M}$} &\colhead{T$_a$}			&  \colhead{$v_{\rm LSR}$}		& \colhead{$\sigma_{\rm obs}$} & \colhead{$\sigma_{\rm nth}$} & \colhead{$\mathcal{M}$}  \\
						& 						& \colhead{(K)} 			&  \colhead{(\kms)} 			& 
\colhead{(\kms)} 			& \colhead{(\kms)}  			& 			 		&  \colhead{(K)} 			&  
\colhead{(\kms)} 			& \colhead{(\kms)}  			& \colhead{(\kms)}		& 					
} 
\decimalcolnumbers
\startdata
G10.99 & 1 & 0.41(0.03) &  29.75(0.03) &  0.28(0.03) &  0.27(0.03) &  1.24(0.14) &  0.33(0.03) &  29.69(0.05) &  0.23(0.12) &  0.21(0.13) &  0.95(0.61) \\
G10.99 & 2 &  ...  &   ...  &   ...  &   ...  &   ...  &  0.57(0.06) &  29.89(0.09) &  0.28(0.25) &  0.27(0.27) &  1.27(1.29) \\
G10.99 & 3 & 0.51(0.05) &  29.63(0.05) &  0.44(0.05) &  0.43(0.05) &  2.07(0.25) &  0.32(0.03) &  29.97(0.13) &  0.53(0.38) &  0.52(0.39) &  2.54(1.88) \\
G10.99 & 4 &  ...  &   ...  &   ...  &   ...  &   ...  &  0.22(0.02) &  30.01(0.10) &  0.14(0.07) &  0.10(0.10) &  0.46(0.43) \\
G10.99 & 5 & 0.79(0.11) &  29.90(0.03) &  0.18(0.03) &  0.16(0.03) &  0.82(0.22) &  0.71(0.07) &  29.72(0.06) &  0.22(0.13) &  0.20(0.15) &  1.01(0.77) \\
G10.99 & 6 &  ...  &   ...  &   ...  &   ...  &   ...  &   ...  &   ...  &   ...  &   ...  &   ...  \\
G10.99 & 7 & 1.74(0.07) &  29.56(0.01) &  0.19(0.01) &  0.17(0.01) &  0.82(0.06) &  0.55(0.05) &  29.41(0.04) &  0.18(0.05) &  0.15(0.05) &  0.73(0.26) \\
G10.99 & 8 & 0.44(0.06) &  29.83(0.05) &  0.30(0.05) &  0.28(0.05) &  1.35(0.26) &   ...  &   ...  &   ...  &   ...  &   ...  \\
G10.99 & 9 & 0.66(0.07) &  29.52(0.04) &  0.35(0.04) &  0.34(0.04) &  1.69(0.23) &  0.56(0.06) &  29.46(0.07) &  0.20(0.08) &  0.17(0.09) &  0.87(0.44) \\
G10.99 & 10 & 0.29(0.05) &  30.46(0.05) &  0.28(0.05) &  0.26(0.06) &  1.23(0.27) &  0.42(0.04) &  30.17(0.05) &  0.11(0.05) &  0.06(0.09) &  0.30(0.44) \\
G10.99 & 11 & 0.87(0.16) &  29.78(0.03) &  0.13(0.02) &  0.09(0.04) &  0.43(0.18) &   ...  &   ...  &   ...  &   ...  &   ...  \\
\enddata
(This table is available in its entirety in a machine-readable form in the online
journal. A portion is shown here for guidance regarding its form and content.)
\tablenotetext{}{Summary of the molecular line parameters derived for all the cores embedded in the 12 IRDCs observed with ALMA. Columns (1) and (2) show the short name of the parental IRDC and dense cores, respectively. 
The \n2dp parameters obtained by the hyperfine line fits are presented in columns (3)--(5), columns (6) and (7) shows the observed velocity dispersion and Mach number. 
Columns (8)--(10) shows the \dcop parameters obtained by the Gaussian fits. The nonthermal velocity dispersion and Mach number are shown in columns (11) and (12), respectively. 
The corresponding uncertainty is given in parentheses. 
Dashes denote no available data. 
}
\end{deluxetable*}
%%%----------------------------------------------------

\subsubsection{Effects of Combining the Single-dish Observations with the Interferometric Data}
We compared the line properties of the spectra obtained 
from the 12m7m datasets with the 12m7mTP datasets. 
We found no difference in the $v_{\rm LSR}$ between 
the spectra for any of the three molecules from the 
present study (\n2dp, \dcop, and C$^{18}$O), whether 
we included or excluded TP observations.

For \dcop, the spectra obtained from the feathered images 
have a velocity dispersion $\sigma_{\rm obs}$ on average 
10\% larger than the spectra obtained from the interferometric 
data alone. In addition, for 63\% of the sample, this 
difference is lower than 10\%.  For the \n2dp data, 
we also see a similar behaviour; however, the 
differences of $\sigma_{\rm obs}$ between 
the feathered and nonfeathered data are on average 
only 3\%. This suggests that the \dcop line 
traces more extended gas than the \n2dp emission 
toward dense cores.

The largest discrepancy was found for the C$^{18}$O 
velocity dispersion, where the differences between the 
interferometric data and the one including the 
single-dish are of the order of $\sim$30\%. 
This is because the C$^{18}$O traces relatively 
lower-density gas compared to the \dcop and \n2dp, 
such as the low-density envelope of  cores 
\citep{Walsh2004,Tychoniec2021}. 

%This is because the C$^{18}$O, in addition to trace 
%the envelope of the cores \citep{Walsh2004}, 
%also traces the more 
%extended and diffuse gas from the clump itself. 

The maximum scale recovered by the 7 m array 
corresponds to $\sim$0.58 pc (30\arcsec at the 
averaged source distance of 4 kpc), which is 
much larger than the detected core typical 
size of  $\sim$0.02 pc. Also, the sizes of the 
cores were determined from their dust continuum 
emission, which only have the emission from the 12  
and 7 m arrays. Therefore to have a better comparison 
between the properties of the cores and their envelopes, 
and to avoid the more extended emission that might be 
associated with the clump itself (intra-clump gas not 
associated with the dense cores), for the rest of our 
analyses, we will use the properties of the line emission 
derived from the data cubes that only contain the 12  
and  7 m arrays.

In addition, we have assessed how much flux is recovered 
in the 12m7m continuum images using the single-dish 
data retrieved from the APEX ATLASGAL survey 
at 870~$\mu$m \citep{schuller-2009}. 
We have scaled the 870~$\mu$m flux to 1.3~mm flux 
assuming a dust emissivity spectral index of $\beta$ = 1.5;  
$F_{\rm 1.3 mm,exp} = F_{\rm 0.87 mm} \, (1.3/0.87)^{-(\beta+2)}$. 
The continuum images recover about 10--33\% 
($F_{\rm 1.3 mm,12m7m}/F_{\rm 1.3 mm,exp}$) of 
single-dish continuum flux, with a mean value of 
21\% \citep[see also][]{Sanhueza-2019}.
%, which is lower than the \dcop and \n2dp line  images. 
For the \dcop and \n2dp lines, their 12m7m 
data can recover about 79\% and 92\% of the 
12m7mTP flux \citep[see also][]{Li-2022b}. 
This suggests that the continuum images 
recover less extended emission than the \dcop 
and \n2dp images made using the 12m7m data.

\subsection{Gas Properties of the Cores}
In general, the emission from the \n2dp and/or \dcop  
traces well the dust continuum core in the ASHES 
observations. For the 294 cores 
detected in the 12 clumps, we detected \dcop emission 
toward 116~cores (40\%) and \n2dp emission toward 
54~cores (18\%). We detected both \dcop and \n2dp 
emission toward 41~cores (14\%). In total, 129~cores 
(44\%) were detected in either \dcop and/or \n2dp.

More than half of dust continuum cores present no 
detectable emission from either \n2dp or \dcop. 
Comparing the physical properties of the cores 
with detected and undetected deuterated species, 
we found that 
the ones without emission from deuterated species have 
lower integrated fluxes in their continuum emission 
($\langle S_{\rm cont,nodeu} \rangle=1.5$ mJy, compared 
to $\langle S_{\rm cont,deu} \rangle=4.4$ mJy), smaller 
masses ($\langle M_{\rm gas,nodeu} \rangle=0.95$ \Mo, 
compared to $\langle M_{\rm gas,deu} \rangle=2.87$ \Mo), 
and lower surface densities 
($\langle \Sigma_{\rm nodeu} \rangle=0.36$ g cm$^{-2}$, compared to 
$\langle \Sigma_{\rm deu} \rangle=0.70$  g cm$^{-2}$).

Since the emission from the deuterated species is rather 
weak, with the signal-to-noise in both species going up 
to 10, the lack of detection of deuterated species could 
be due to either the sensitivity of our observations or 
their abundances being too low in the gas phase, or both. 
Given that we want to compare the properties of the gas 
and the dust, we only focus on those cores that show 
emission in both the dust continuum and molecular lines.

The average values of the core line velocity dispersion 
($\sigma_{\rm obs}$) are similar in both molecules, with 
$\sigma_{\rm obs,N_2D^+}=0.23$ \kms and 
$\sigma_{\rm obs,DCO^+}=0.32$ \kms.
For the 41~cores that showed emission in both molecules, 
we compared the observed velocity dispersion 
($\sigma_{\rm obs}$) and the central velocity 
($v_{\rm LSR}$). Figure~\ref{fig:diff} shows the 
differences of these two quantities between \dcop and \n2dp.  
In general, the central velocity  differences have a 
normal distribution centered around 0.05 \kms. 
The \dcop lines are slightly broader than \n2dp lines, 
which could be due to unresolved hyperfine structure.  
The velocity dispersion of \dcop is derived from the 
single Gaussian fit, whereas \n2dp is derived from 
the hfs fit.  In two cases (these two cores are not shown in 
Figure~\ref{fig:diff}), the central velocity differences 
are up to 2.10 \kms (G14.49-\#6) and 0.77 \kms 
(G14.49-\#11).  
This is most likely because their local environments have been  
significantly affected by the nearby strong protostellar 
outflow \citep[see Figure~1 of][]{Li-2022b}.

%----------------------------------------------------
\begin{figure*}[!ht]
\centering
\includegraphics[scale=0.24]{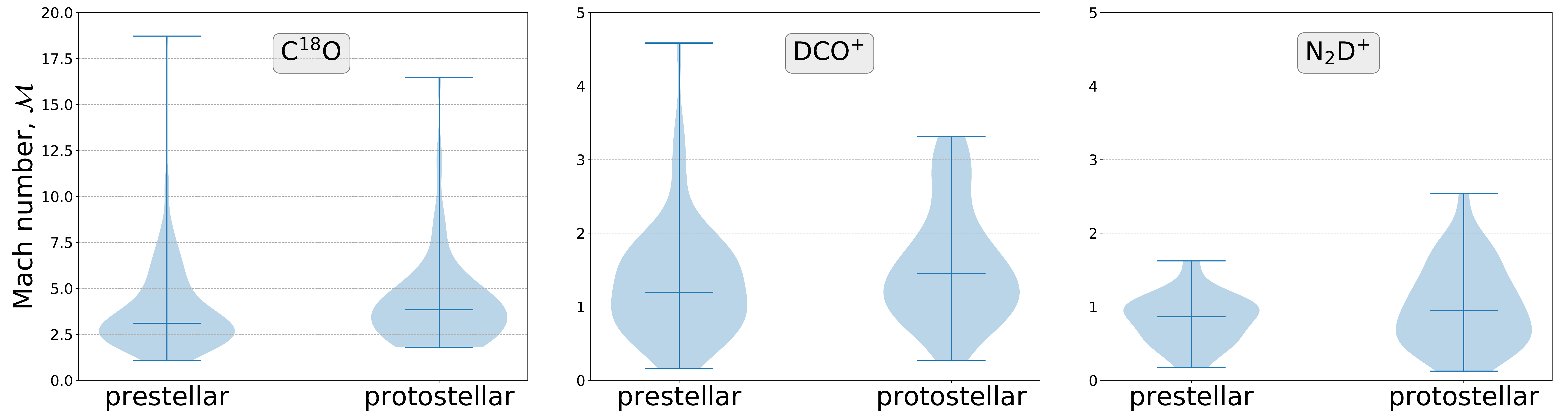}
\caption{
Violin plots of the $\mathcal{M}$ distributions for each line. 
The shape of each distribution shows the probability density 
of the data smoothed by a kernel density estimator. 
The blue horizontal bars from the top to bottom in each violin 
plot represent the maximum, mean, and minimum values, 
respectively. 
}
\label{fig:sig}
\end{figure*}
%%----------------------------------------------------

We found that 71\% of the cores show a difference in 
the \dcop and \n2dp central velocity $<0.17$ \kms, 
which is smaller than the spectral resolution of 
our data, and 94\% of all cores shows a difference 
$<0.34$ \kms. 
Thus, the emission from these two molecules is 
likely arising from the same physical location 
in the cores. Therefore, for the analysis of 
this paper, we will use the velocity dispersion 
obtained from the \dcop emission, and in the 
case that \dcop is not detected, we will use 
the velocity dispersion obtained from the \n2dp 
line. 
The measured velocity dispersion have a mean value 
$\langle \sigma_{\rm obs} \rangle=0.32$ \kms, 
ranging between 0.1 and 0.77 \kms. 
The nonthermal velocity dispersion is given by 
$\sigma_{\rm nt}^2=\sigma_{\rm obs,int}^2 - 
\sigma_{\rm th,m}^2$, where $\sigma_{\rm obs,int}$ 
is the intrinsic observed 
velocity dispersion after removing the smearing 
effect due to the channel width using 
$\sigma_{\rm obs,int} = 
\sqrt{(\rm FWHM_{\rm obs}^2 - FWHM_{\rm ch}^2)/8ln2}$ 
= $\sqrt{\sigma_{\rm obs}^2 - \rm FWHM_{\rm ch}^2/8ln2}$, 
$\sigma_{\rm obs}$ is the observed velocity dispersion, 
FWHM$_{\rm ch}$ is the channel width, 
and $\sigma_{\rm th,m}= (k_{\rm B}T/\mu m_H)^{1/2} 
= 9.08 \times 10^{-2} \, {\rm km \,s^{-1}} \rm 
\left(\frac{T}{K} \right)^{0.5} \mu^{-0.5}$ 
is the thermal velocity dispersion. 
The value of $\mu$ 
for both \dcop and \n2dp is 30.  
The sound speed $c_{\rm s}$ can be estimated using 
a mean molecular weight per free particle of 
$\mu_{\rm p}$ = 2.37, which assumes a typical 
interstellar abundance of H, He, and metals 
\citep{2008A&A...487..993K}. 
For the cores, $\sigma_{\rm th,m}$ ranges between 
0.04 and 0.08 \kms, with a mean value of 0.06 \kms, 
and $\sigma_{\rm nt}$ ranges between 0.03 and 
0.77 \kms, with a mean value 
$\sigma_{\rm nt}=0.27$ \kms.

\subsection{Gas properties of the envelopes}
The C$^{18}$O emission is extended and filamentary 
in the clumps. In our sample, 91\% of the cores have 
C$^{18}$O emission, and we speculate that C$^{18}$O is 
likely depleted in the cores where there is no 
detection \citep{Sabatini-2022}.  
%There is a fraction  (\new{XX\%}) of dense core present 
%distinct separate double peak profiles in the C$^{18}$O 
%emission.
Using the core systemic velocity defined 
by other dense gas tracers (i.e., \n2dp, \dcop, \h2co, 
or \ch3oh), we were able to fit a Gaussian to the 
C$^{18}$O line profiles to \cocores~out of the 
\ndustcores~cores. 
C$^{18}$O was observed in a spectral window with 
coarser spectral resolution (1.3 \kms), which is 
in general still narrower than the derived typical FWHM 
(2.2 \kms)  for this transition. However, the coarser 
spectral resolution might make us overestimate  
the true internal velocity dispersion of the gas 
(inside cores). For these 
reasons, we focused our analysis on the deuterated 
species, and used the parameters derived from the 
C$^{18}$O only as a reference, to compare the 
properties of the core's gas with the properties of 
their envelopes. 
The velocity dispersions for the C$^{18}$O emission, 
as expected, have larger values than the ones found 
for the deuterated species, ranging from 
$\sigma_{\rm obs,C^{18}O}=0.4$ to 3.4 \kms, 
with a mean value of 0.9 \kms.

\subsection{Virial mass and virial parameter}
\label{sec:virial}
To determine the gravitational state of the cores, we 
calculated their virial parameter 
$\alpha=M_{\rm vir}/M_{\rm gas}$, where 
$M_{\rm vir}$ is the virial mass, and $M_{\rm gas}$ 
is the mass derived from the dust continuum emission. 
The virial parameter, $\alpha$, is 
commonly used to assess the gravitational state of 
molecular clouds. A value of $\alpha<1$ suggests 
that a core is gravitationally unstable to collapse, 
and turbulence alone cannot maintain its stability 
\citep{Bertoldi-1992}. 
Nonmagnetized cores with $\alpha \sim 1$  and $\alpha<2$ 
are considered to be in hydrostatic equilibrium and 
gravitationally bound, respectively.  A value of $\alpha>2$, 
on the other hand, suggests that the core is gravitationally 
unbound \citep{Kauffmann2013}, and if we ignore 
additional support mechanisms  
such as the magnetic field or external pressure, then the 
core might be a transient object.

The virial mass was calculated using the observation  
measurements following \citep{Bertoldi-1992}:
\begin{equation}
M_{\rm vir}= \frac{5}{a\beta} \frac{\sigma_{\rm tot}^2 R}{G},
\end{equation}
\noindent where $R$ is the core radius, 
$a = (1-b/3)/(1-2b/5)$ is the correction factor for a 
power-law density profile $\rho \propto R^{-b}$, 
$\beta = ({\rm arcsin}\,e)/e$ is the geometry factor 
\citep[see][for detailed 
derivation]{1983AJ.....88.1626F,2013ApJ...768L...5L}, 
$\sigma_{\rm tot}$ is the total velocity dispersion of 
the gas in the core, and $G$ is the gravitational constant.  
The eccentricity $e =  \sqrt{1 - f_{\rm int}^{2}}$ is 
calculated by the intrinsic axis ratio, $f_{\rm int}$, 
of the dense cores.  
The $f_{\rm int}$ can be estimated from observed 
axis ratio, $f_{\rm obs}$, with 
$f_{\rm int}  = \frac{2}{\pi} \,f_{\rm obs} \, \mathcal{F}_{1}(0.5, 0.5, -0.5, 1.5, 1, 1-f_{\rm obs}^{2})$ \citep{1983AJ.....88.1626F}, 
where $\mathcal{F}_{1}$ is the Appell 
hypergeometric function of the first kind. 
For the dense cores, the derived $\beta$ ranges from 
1.0 to 1.4, with a mean value of 1.2. 
Here we adopted a typical density profile index 
$b = 1.6$ for all dense cores 
\citep[e.g.,][]{2002ApJ...566..945B,2012ApJ...754....5B,
2014ApJ...785...42P,Li-2019b}. 
The velocity dispersion is given by 
$\sigma_{\rm tot}^2 = \sigma_{\rm nt}^2 + c_{\rm s}^2$, 
and it reflects a combination between the nonthermal 
motions of the gas within the core, $\sigma_{\rm nt}$, 
and the thermal motion of the particle mean mass, 
$c_{\rm s}^2$.  $c_{\rm s}^2$ ranges 
between 0.16 and 0.28 \kms, with a mean value of 
0.23 \kms.

We found that the values of the virial masses range 
from 0.22 to 8.61 \Mo, and the mean value is 
1.89~\Mo for all the \totalmolcores~cores analyzed. 
The virial parameter $\alpha=M_{\rm vir}/M_{\rm gas}$ 
ranges from 0.15 to 8.91, with a mean value of 1.39. 
Seventy-three out of  \totalmolcores~cores have 
virial parameters  below 1,  28 cores have virial 
parameters between 1 and 2, and 28 cores have 
$\alpha$ larger than 2. 
The derived values of the virial mass and virial 
parameter are summarized in Table~\ref{table_cores}.

\begin{table*}
\scriptsize
% \tiny
\centering
\caption{Median and mean values of derived core properties for each category}
\label{tab:pcore}
\begin{tabular}{l c c c c c c c c c  c c c}
\toprule
Name & 
\multicolumn{2}{c}{Prestellar} & 
\multicolumn{8}{c}{Protostellar} &
\multicolumn{2}{c}{All} \\ 
\cmidrule(l{2pt}r{2pt}){4-11}
	 & & &
\multicolumn{2}{c}{Outflow Core} &
\multicolumn{2}{c}{Warm Core} &
\multicolumn{2}{c}{Warm\&outflow Core} &
\multicolumn{2}{c}{Sum$^{a}$} & \\
\cmidrule(l{1pt}r{2pt}){2-3} \cmidrule(l{2pt}r{2pt}){4-5} 
\cmidrule(l{1pt}r{2pt}){6-7} \cmidrule(l{2pt}r{2pt}){8-9}
\cmidrule(l{1pt}r{2pt}){10-11} \cmidrule(l{2pt}r{2pt}){12-13} 
& Median & Mean 
& Median & Mean 
& Median & Mean 
& Median & Mean 
& Median & Mean 
& Median & Mean     \\   
	\midrule    
$T_{\rm NH_{3}}$ & 14.1 & 14.2 & 12.8 & 13.8 & 14.8 & 14.3 & 15.3 & 14.9 & 14.8 & 14.4 & 14.4 & 14.3	\\ \hline 
$M$ & 0.63 & 1.15 & 1.26 & 2.18 & 0.69 & 2.00 & 3.11 & 4.66 & 1.41 & 3.11 & 0.77 & 1.79	\\ \hline 
$n_{\rm H_2}$  & 1.11  & 1.58  & 1.59  & 2.62  & 1.26  & 2.82  & 3.59  & 5.55  & 2.40  & 3.87  & 1.32  & 2.34 	\\ \hline
$N_{\rm peak}$(H$_{2}$) & 3.01  & 3.58  & 5.40  & 7.26  & 3.35  & 5.55  & 11.40  & 15.57  & 6.32  & 9.95  & 3.58  & 3.73 	\\ \hline 
$\Sigma$ & 0.29  & 0.34  & 0.46  & 0.60  & 0.30  & 0.54  & 0.96  & 1.27  & 0.64  & 0.85  & 0.35  & 0.51 	\\ \hline
$R$ & 0.012  & 0.014  & 0.014  & 0.015  & 0.012  & 0.013  & 0.013  & 0.015  & 0.013  & 0.014  & 0.013  & 0.014 	\\ \hline
$M_{\rm vir}$ & 1.05  & 1.67  & 1.31  & 1.33  & 1.28  & 2.29  & 2.00  & 2.42  & 1.58  & 2.18  & 1.25  & 1.89 	\\ \hline
$\alpha$ & 1.01  & 1.54  & 0.52  & 0.61  & 1.88  & 2.35  & 0.40  & 0.53  & 0.58  & 1.19  & 0.87  & 1.39 	\\ \hline
$\mathcal{M}_{\rm  N_{2}D^{+}}$ & 0.9  & 0.8  & 1.0  & 1.1  & 0.8  & 0.9  & 1.0  & 1.0  & 0.9  & 1.0  & 0.9  & 1.0 	\\ \hline
$\mathcal{M}_{\rm DCO^{+}}$ & 1.2  & 1.3  & 1.4  & 1.3  & 1.6  & 1.7  & 1.5  & 1.5  & 1.5  & 1.6  & 1.3  & 1.4 	\\ \hline
$\mathcal{M}_{\rm C^{18}O}$ & 3.1  & 3.8  & 3.8  & 4.2  & 4.0  & 4.4  & 3.6  & 4.0  & 3.8  & 4.2  & 3.3  & 3.9 	\\
\bottomrule
\end{tabular}
\tablenotetext{}{Notes. 
The unit for $M_{\rm gas}$, $n_{\rm H_2}$,  
$N_{\rm peak}$ (H$_2$), $\Sigma$, $R$, and $M_{\rm vir}$ are 
$M_{\odot}$, $\times 10^{6}$ cm$^{-3}$, $\times 10^{22}$ cm$^{-2}$, g cm$^{-2}$, pc, 
and $M_{\odot}$, respectively. 
}
\tablenotetext{a}{All protostellar cores include outflow core, warm core, and warm and outflow core.}
\end{table*}

%%-------------------------------------------------------
\begin{figure*}
\centering
\includegraphics[trim={0cm 0cm 0cm 0cm}, 
clip,width=0.75\textwidth]{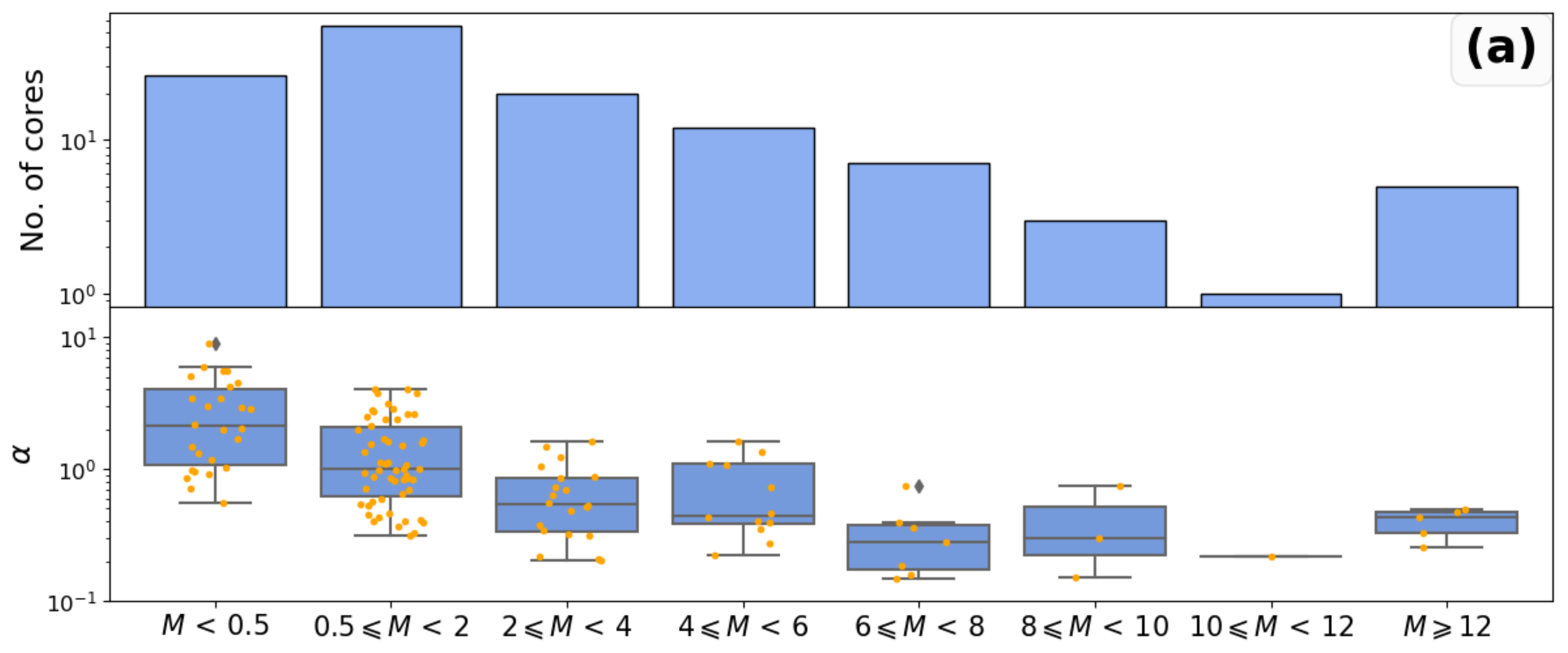}\\
\includegraphics[trim={0cm 0cm 0cm -1.5cm}, clip,width=0.95\textwidth]{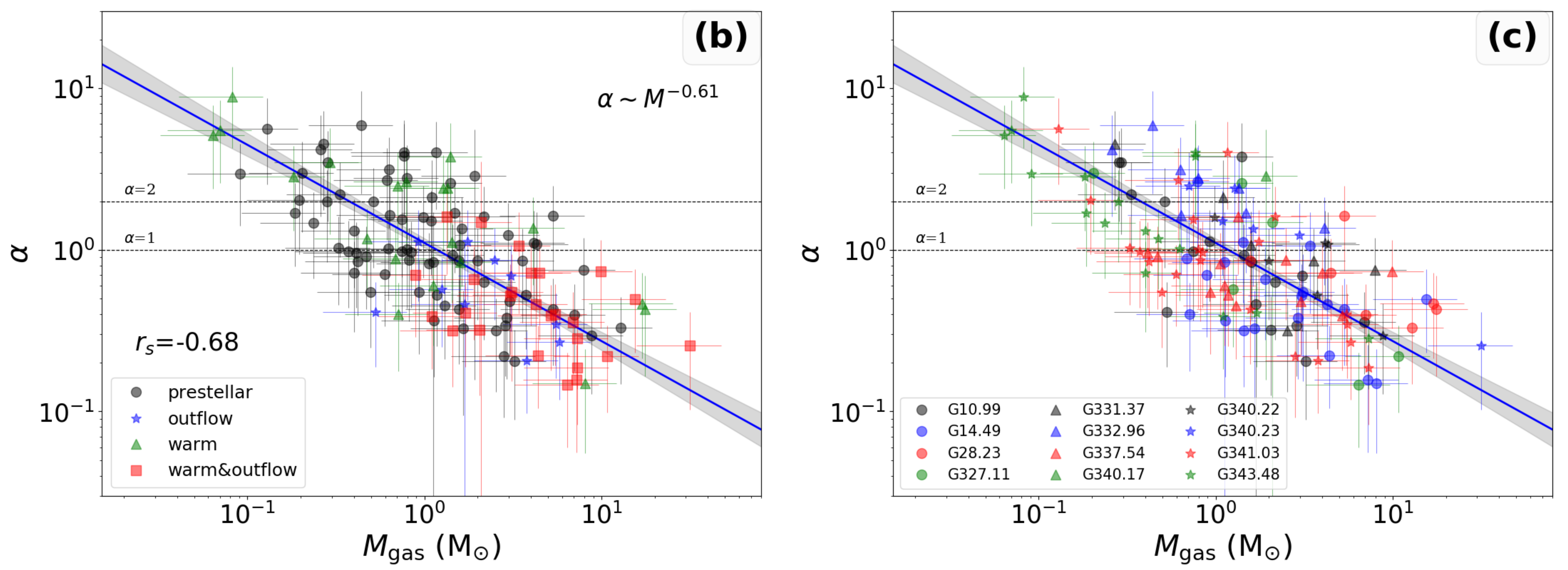}
\caption{
Panel (a): upper panel shows the number of 
sources per mass bin, and the bottom panel shows 
a box plot of the values of the virial parameter per bin. 
In the box plot, the mean value of the virial parameter  
is shown with a horizontal line. 
Panels (b) and (c): $\alpha$ versus mass for all the 
cores where molecular emission from either \n2dp 
or \dcop was detected. Panel (b) shows the 
cores color coded by their evolutionary stage 
classification \citep{Li-2022b}, and panel (c) shows the 
cores color coded by their parental molecular cloud. 
The solid line in both figures shows the linear
regression to all the point in the plot, gives a slope 
of -0.61 $\pm$ 0.06, and the gray shadowed area 
shows the 1$\sigma$ confidence interval 
for the fit. The Spearman's rank test returns a 
coefficient of $r_{s}$ = -0.68. 
} \label{fig:alpha}%
\end{figure*}
%%-------------------------------------------------------

\subsection{Mach Number}
\label{sec:mach}
To analyze the amount of turbulence within the cores, 
we calculated the one-dimensional (1D) turbulent Mach 
number ($\mathcal{M} = \sigma_{\rm nt}/c_{\rm s}$).  
The $\mathcal{M}$ shows no significant difference 
between the prestellar and protostellar cores for \co18,  
\dcop, and \n2dp (see Figure~\ref{fig:sig}), 
indicating that the nonthermal motions 
of the molecular gas traced by \co18, \dcop, 
or \n2dp have not yet been significantly influenced 
by protostellar activity.   
%For \n2dp, $\mathcal{M}$ in the protostellar cores is 
%relatively larger than that in the prestellar ones.  
%These result suggests that \n2dp is more sensitive to 
%the protostellar activity than \dcop, in terms of line width. 
There are 98\% (53/54) and 82\% 
(95/116) of $\mathcal{M}$ derived from 
\n2dp and \dcop smaller than 2, respectively. 
\n2dp and \dcop are typical tracers of cold and 
dense gas that are closely related to core formation. 
The transonic $\mathcal{M}$ suggests that the  
cold and dense gas in the cores have been so far 
weakly affected by protostellar feedback 
\citep[see also][]{Li-2020b}, albeit a small fraction 
(14\%) of embedded cores is associated with outflow 
activity. Therefore, these 70 \um dark clumps are 
ideal targets for investigating the initial conditions of 
massive star and cluster formation.

Overall,  $\sigma_{\rm obs}$ and $\mathcal{M}$ of \co18  
are systemically higher  than  \n2dp 
and \dcop. This is because the latter two deuterated 
species are preferentially tracing  cold and dense gas,  
whereas \co18 probes relatively less dense gas.

For all cores,  the $\mathcal{M}$ derived from \dcop 
or \n2dp ranges from 0.1 to 4.6, with the mean and 
median values of 1.4 and 1.3, respectively.  We found 
that 44 cores show subsonic motions 
($\mathcal{M} \leqslant $1),  
and 64 cores present transonic motions 
(1$< \mathcal{M} \leqslant $2). 
This suggests that the motions in most of the 
deuterated  cores (108/129 = 84\%) are subsonic 
or transonic dominated.

\section{Discussion}
\label{sec:dis}
\subsection{Core mass growth}
\label{sec:growth}
As shown in Figure~\ref{fig:NH2}, the protostellar 
cores have higher column densities, gas masses, 
volume densities, and  surface densities than 
those from prestellar cores (see also 
Table~\ref{tab:pcore}). A similar trend is found 
in the individual clumps,  except for G331.37, 
G340.17, and G340.22. The latter two clumps 
are still at very early evolutionary phases 
as evidenced by the nondetection of molecular 
outflows. The G331.37 hosts 1 relatively 
massive (7.96~\Mo) prestellar core, while 
the rest of the prestellar and protostellar 
cores have comparable mass ($<$ 4 \Mo).

These results suggest that the protostellar cores 
are more massive and denser than the prestellar 
cores,  indicating a core mass growth from the 
prestellar to protostellar stages 
\citep[e.g.,][]{2021ApJ...912..156K,Takemura-2022,Nony2023}. 
%Protostellar cores are expected to become more massive 
%and denser than their younger siblings of prestellar 
%cores through accretion of significant material from 
%the natal molecular clump.  
In addition,  
\cite{Li-2020b} found that the more massive cores 
have a longer accretion history than the less 
massive cores. A Kolmogorov-Smirnov (KS) test, 
which returns a probability 
($p$-value) of  two samples being drawn from the same 
population, was used to compare the 
peak column density, gas mass, volume density,  
and surface density of the 
prestellar cores with those of the protostellar cores. 
If the $p$-value is much smaller than 0.05, we can reject 
the null hypothesis that the two samples are drawn from 
the same parent distribution 
\citep{TEEGAVARAPU20191}.  
The KS test reveals that the $p$-value is 
2.5$\times$10$^{-13}$, 1.6$\times$10$^{-6}$, 
1.6$\times$10$^{-6}$, and 6.8$\times$10$^{-12}$  
for peak column density, gas mass, volume density,  
and surface density, respectively, indicating that 
the prestellar cores and the protostellar cores 
in our sample are drawn from significantly 
different populations ($p$-value $\ll$~0.05).

Table~\ref{tab:pcore} also shows that  
warm and outflow cores tend to be more massive 
and denser than the remaining type of cores.  
The warm and outflow cores associated with 
molecular outflows and warm line(s) emission 
could be considered as the most evolved objects 
among the identified cores. This further 
supports that the cores have grown in mass and 
density over  time.

\subsection{Stability}
\label{sec:stab}
Several observational biases could lead to underestimate  
the virial parameter. For instance, the observed line 
width could be underestimated when a particular 
molecular line preferentially traces molecular gas with 
densities above the critical density of the line transition, 
resulting in an underestimation of the virial parameter 
\citep{Traficante-2018b}. In our case, this effect is likely 
insignificant owing to the critical density of the  
transition lines used in this study typically being 
1-2 orders of magnitude larger  
than the effective excitation density because 
of the effect of radiative trapping \citep{Shirley-2015};    
the critical density is $\sim$10$^6$ cm$^{-3}$ 
for \dcop and \n2dp. 
Therefore,  both \dcop and \n2dp can effectively 
trace dense gas within the cores. 
On the other hand, the virial parameter could be 
overestimated if gas bulk motions and background 
emission are not properly considered in the 
calculation of the velocity dispersion 
\citep[e.g.,][]{Singh-2021}. 
We computed the velocity dispersion from the core-averaged 
spectrum for each core, bulk motions (if present) already 
included as they could increase the velocity dispersion 
over the core. 
In addition, the interferometric observations 
of dense gas tracers 
filter out the large-scale emission and, therefore, are 
barely affected by the contamination of background emission. 
Overall, the above-mentioned observational biases do not 
significantly affect  our results.

%%-------------------------------------------------------
\begin{figure*}
\centering
\includegraphics[width=1\textwidth]{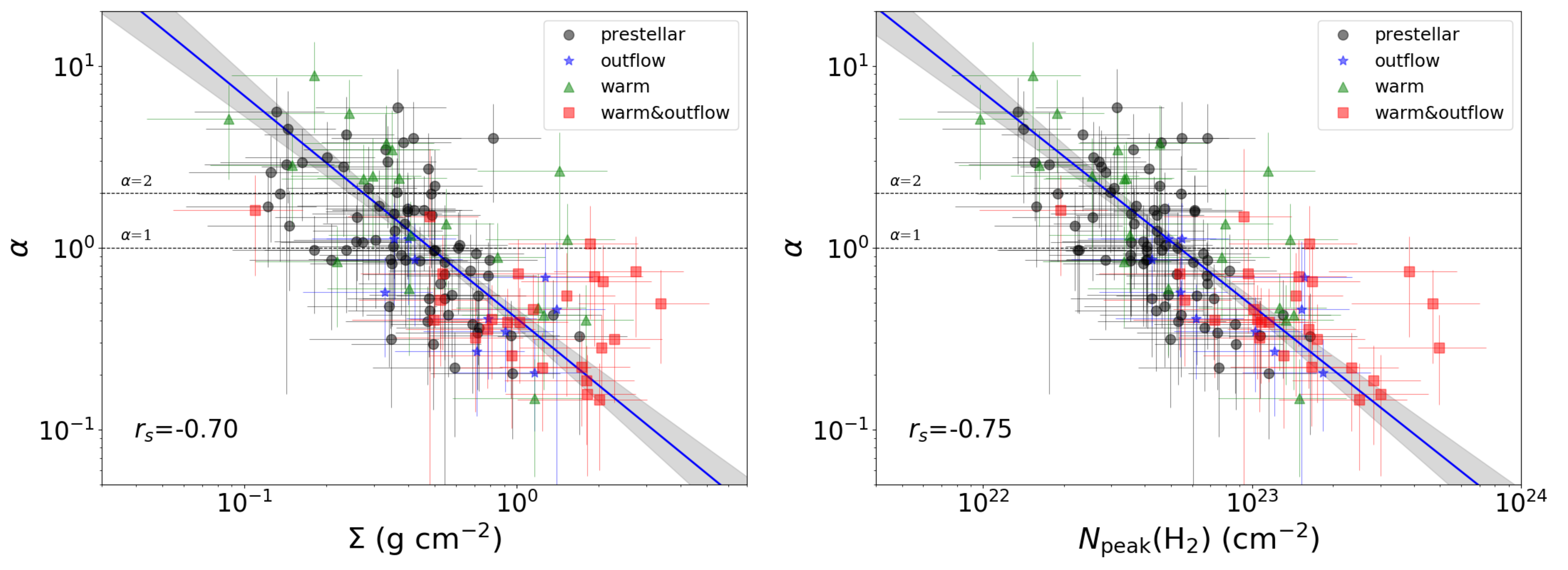}
\caption{Left panel: virial parameter versus surface 
density.
Right panel: virial parameter versus peak column 
density.
The blue solid line in both figures shows the linear
regression to all the point in the plot, and the gray 
shadowed area shows the 1$\sigma$ confidence interval 
for the fit. The best fit gives a slope 
-1.22 $\pm$ 0.15 and -1.17 $\pm$ 0.13 for 
$\alpha$--$\Sigma$ and $\alpha$--$N_{\rm H_2}$, 
respectively. 
The Spearman's rank coefficients are shown in the 
lower left of each panel.
} \label{fig:alpha1}%
\end{figure*}
%%-------------------------------------------------------

\subsubsection{Are massive cores more unstable?}
Comparing the virial parameter of the cores with 
their masses, we find a correlation 
between these two quantities. The value of $\alpha$ 
appears to decrease with increasing core mass  
(see Figure \ref{fig:alpha}), with a Spearman's rank 
correlation\footnote{Spearman rank correlation test 
is a nonparametric measure of the monotonicity of 
the relationship between two variables. 
The correlation coefficient 
$|r_{s}|\geqslant 0.5$ means 
strong correlation, 
$0.5 > |r_{s}|\geqslant 0.3$ means 
moderate correlation, 
$0.3 > |r_{s}|\geqslant 0.1$ means 
weak correlation, and $0.1 > |r_{s}|$ 
means no correlation \citep{Cohen1988}.
If the $p$-value of the correlation is not  
less than 0.05, the correlation is not 
statistically significant.}  
coefficient of $r_{s}$ = -0.68 
($p$-value=1.7$\times10^{-18}$). 
Considering that the virial parameter is 
mathematically inversely proportional to the 
cores' mass, a slope of -1 in a log--log plot 
would suggest that the trend seen is not 
physically meaningful. 
A linear regression\footnote{We fit a
linear regression mode to the data 
using the 
\href{https://linmix.readthedocs.io/en/latest/index.html} 
{LINMIX} \citep{Kelly2007}, 
which is a hierarchical Bayesian 
linear regression routine that 
accounts for errors on both variables. 
The LINMIX is used for the linear 
regression fitting throughout the paper. 
} 
between the $\log{\alpha}$ and 
$\log{M_{\rm gas}}$, gives a slope of 
-0.61 $\pm$ 0.06, indicating that the trend between 
the stability of the cores and their mass is real, 
and not due to an observational bias given by the 
correlation between the two core properties. 
The observed trend of $\alpha$ and 
$M_{\rm gas}$, $\alpha \sim M_{\rm gas}^{-0.61\pm0.06}$, 
is consistent with the expectations for 
pressure-confined cores, in which $\alpha$ 
should depend on mass as 
$\alpha \sim M_{\rm gas}^{-2/3}$ 
\cite[e.g.,][]{Bertoldi-1992}. 
The $\log{\alpha} - \log{M_{\rm gas}}$ correlation  
indicates that the most massive cores tend to be more 
gravitationally unstable. In contrast, the less massive 
cores are gravitationally unbound and may eventually 
disperse if no other mechanism(s) can  help to 
confine them, such as external pressure \citep{Li-2020a}. 
A similar inverse relationship between the virial parameter 
and the core mass has also been seen in other 
IRDCs \citep[e.g.,][]{Li-2019b,Li-2020a}.

As shown in Figure~\ref{fig:alpha1},  there is a strong  
inverse relationship between the virial parameter and 
the core surface density, with a Spearman's rank 
coefficient of $r_{s}$ = -0.70 
($p$-value=6.4$\times10^{-20}$), 
and the best fit 
returns a slope of -1.22 $\pm$ 0.15. This indicates 
that higher surface density cores appear to be more 
gravitationally unstable.  
We also find a strong trend of virial parameters 
and column density, with the higher column density 
tending to show a lower virial parameter 
(see Figure~\ref{fig:alpha1}). 
The Spearman's rank test returns a coefficient of 
$r_{s}$ = -0.75 ($p$-value=1.8$\times10^{-24}$), 
and the best fit gives a slope of 
-1.17 $\pm$ 0.13. The inverse $\alpha-N_{\rm H_2}$ 
relation suggests that the higher-density 
cores tend to be more gravitationally unstable. 
Overall, these results suggest that the more 
massive and denser cores tend to be more 
gravitationally unstable.

Given the angular resolution of our observations, 
it is possible that with observations at higher angular 
resolution the most massive cores could further 
fragment into smaller objects. Assuming that the cores 
are not fragmented, and ignoring the effect of additional 
support (such as magnetic fields), we suggest that cores 
with masses M$_{\rm core}>4$ $M_\odot$ in each clump 
are strongly self-gravitating 
($\langle \alpha \rangle \, = 0.37 < 1$) 
and prone to collapse, except for core 
\#8-G28.27, \#1-G332.96, and \#4-G340.22 that 
have $\alpha$ of 1.6, 1.4, and 1.1, respectively.

The most massive core in all the 12 IRDCs is core \#1
in G340.23, with a mass of  31.66 $M_\odot$. For the 
remaining IRDCs, the  most massive cores have masses of 
$\sim$10~$M_\odot$. If these cores will form high-mass 
stars, then while collapsing they might still be accreting 
material from their environment in order to have enough 
mass ($>27$ $M_\odot$, assuming a star formation 
efficiency of 30\%) to form a $>$8 $M_\odot$ star. 
Indeed, this is happening for one of the cores of G331.37 
(core \#1), where additional observations of the HCO$^+$ 
molecular gas show that this subvirialized core is 
accreting material at a high rate from its environment 
\citep{Contreras-2018}. If this core in G331.37 continues 
to accrete at its current rate 
($2.0 \times 10^{-3}$ $M_\odot$ yr$^{-1}$), it is expected 
that in a freefall time (3.3 $\times$ 10$^{4}$ yr) the 
core will have 82 $M_\odot$, gathering a large mass 
reservoir and allowing the formation of high-mass stars. 
Although this is only one example, we may expect 
similar accretion rates in several ASHES cores given 
their low virial parameters and massive gas reservoir 
in their natal clumps (see Table~\ref{summaryclumps}). 
In addition, the filamentary structures connected to cores 
can also  transport material from the parental clump 
onto cores and thus further increase their masses. 
For example, the mass flow rate along a  
filamentary structure feeding a protostellar core 
in one of the ASHES targets, G14.49, is about 
$2.2 \times 10^{-4}$ $M_\odot$ yr$^{-1}$ 
\citep{Redaelli2022}. Such accretion rates along 
filamentary structures are not unusual in high-mass 
star-forming regions \citep[e.g.,][]{Sanhueza-2021}.

\subsubsection{Does core stability change with the evolutionary stage of cores?}
\label{stability-cores}
Prestellar and protostellar cores have a mean virial 
mass of $\langle M_{\rm vir} \rangle=1.67$~\Mo and 
$\langle M_{\rm vir} \rangle=2.18$~\Mo, respectively. 
Overall, protostellar cores have higher virial mass 
than that of prestellar cores, indicating that the virial 
mass tends to increase with the evolutionary stage  
of cores. 
This is likely because the velocity dispersion 
toward protostellar cores is slightly larger than that 
of prestellar cores; the mean velocity 
dispersion derived from \n2dp or \dcop 
is 0.28 and 0.32 \kms for prestellar 
and protostellar cores, respectively.  On the other hand, 
the virial parameter of protostellar cores 
is slightly lower than the prestellar cores as shown in 
Figure~\ref{fig:alpha}. The prestellar cores have a mean 
virial parameter of $\langle \alpha \rangle=1.5$, 
while the protostellar cores have a mean value of 
$\langle \alpha \rangle=1.2$. 
This is because the protostellar cores have relatively 
higher gas mass than the prestellar cores. 
The  lower value of the virial parameter, ignoring 
additional forces of support, for protostellar cores is in 
agreement with the scenario where protostars might 
be embedded in them, and thus the gravitational collapse 
has already started in these regions. 
Figures \ref{fig:alpha} and \ref{fig:alpha1} suggest that 
the virial parameter decreases with mass and density.   
In Section~\ref{sec:growth}, we also find that 
both mass and density of cores increase from the 
prestellar to the protostellar phase. These observational 
results indicate that the virial parameter tends to 
decrease with the evolutionary stage of cores, and 
therefore cores become more gravitationally 
unstable with evolution.

%%-----------------------------------------------
\begin{figure*}
\centering
\includegraphics[width=0.45\textwidth]{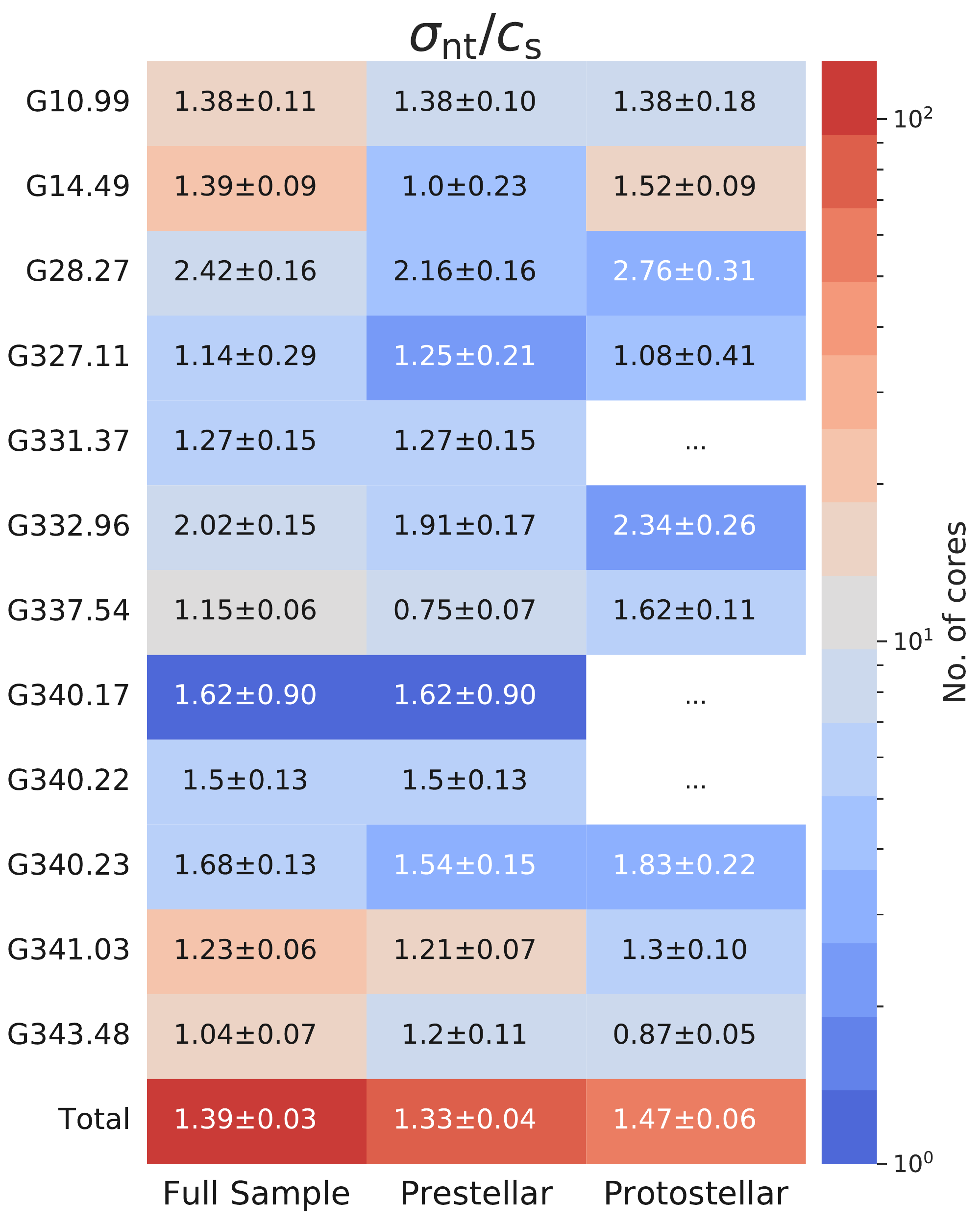}
\includegraphics[width=0.45\textwidth]{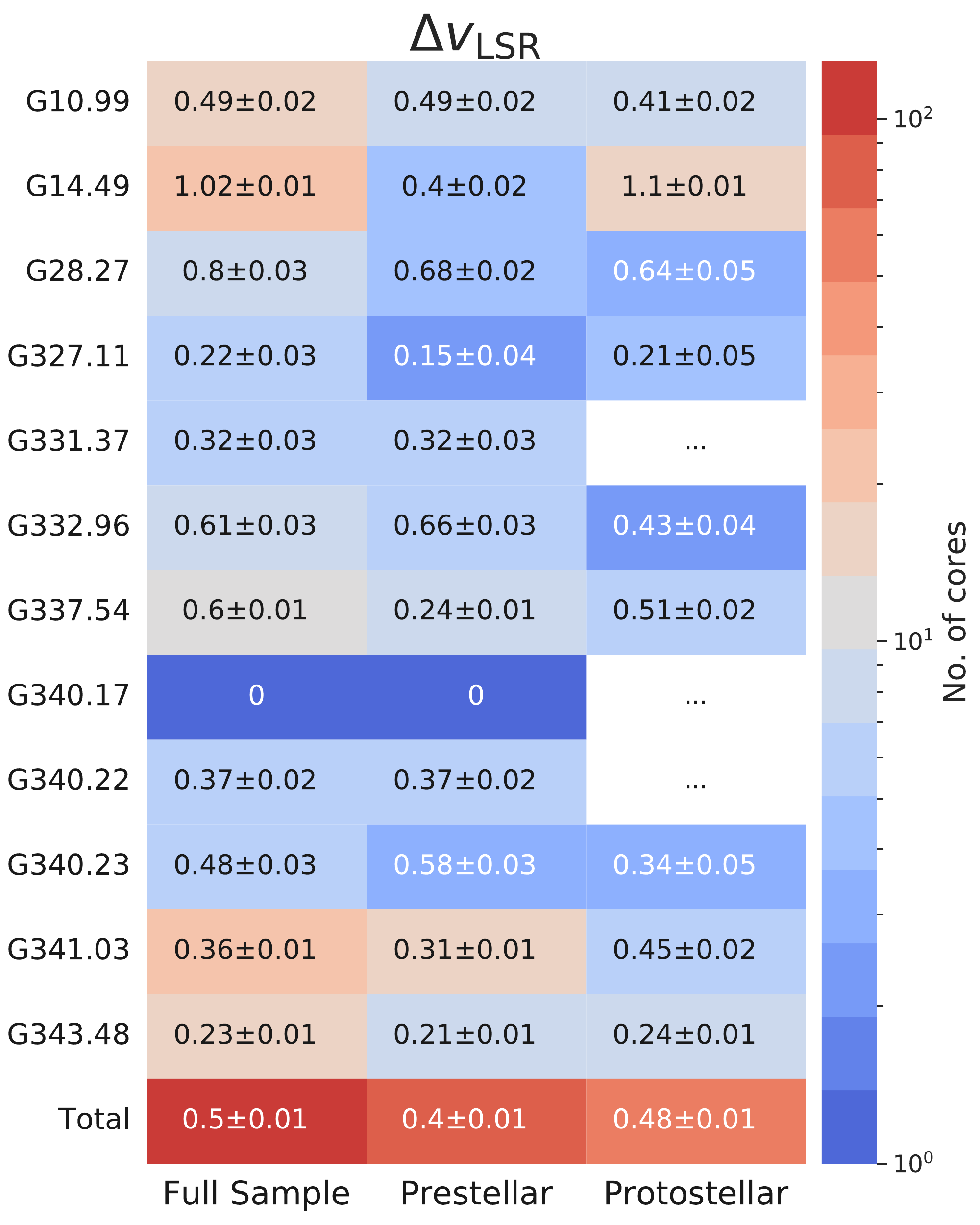}
\caption{Left panel: the Mach number 
($\sigma_{\rm nt}/c_{\rm s}$) of the 
cores, obtained from \dcop or \n2dp 
molecular line emission for each clump. The total 
row shows the mean value of the velocity dispersion 
for the clumps that have a measurement in both the 
prestellar and protostellar sample. 
Right panel: core-to-core velocity dispersion of the 
cores $v_{\rm LSR}$ obtained from \dcop or  
\n2dp molecular line emission for each clump. 
The total row shows the mean value of the core-to-core 
velocity dispersion for the clumps that have a 
measurement in both the prestellar and 
protostellar sample. The color scale shows 
the number of cores considered to determine the velocity 
dispersion. `0' means only one available core.
Dashes denote no available core.} 
\label{fig:hmap}%
 \end{figure*}
%%-----------------------------------------------

%%------------------------------------------------
\begin{figure*}
\centering
\includegraphics[trim={0cm 0cm 0cm 0cm}, 
clip,width=0.75\textwidth]{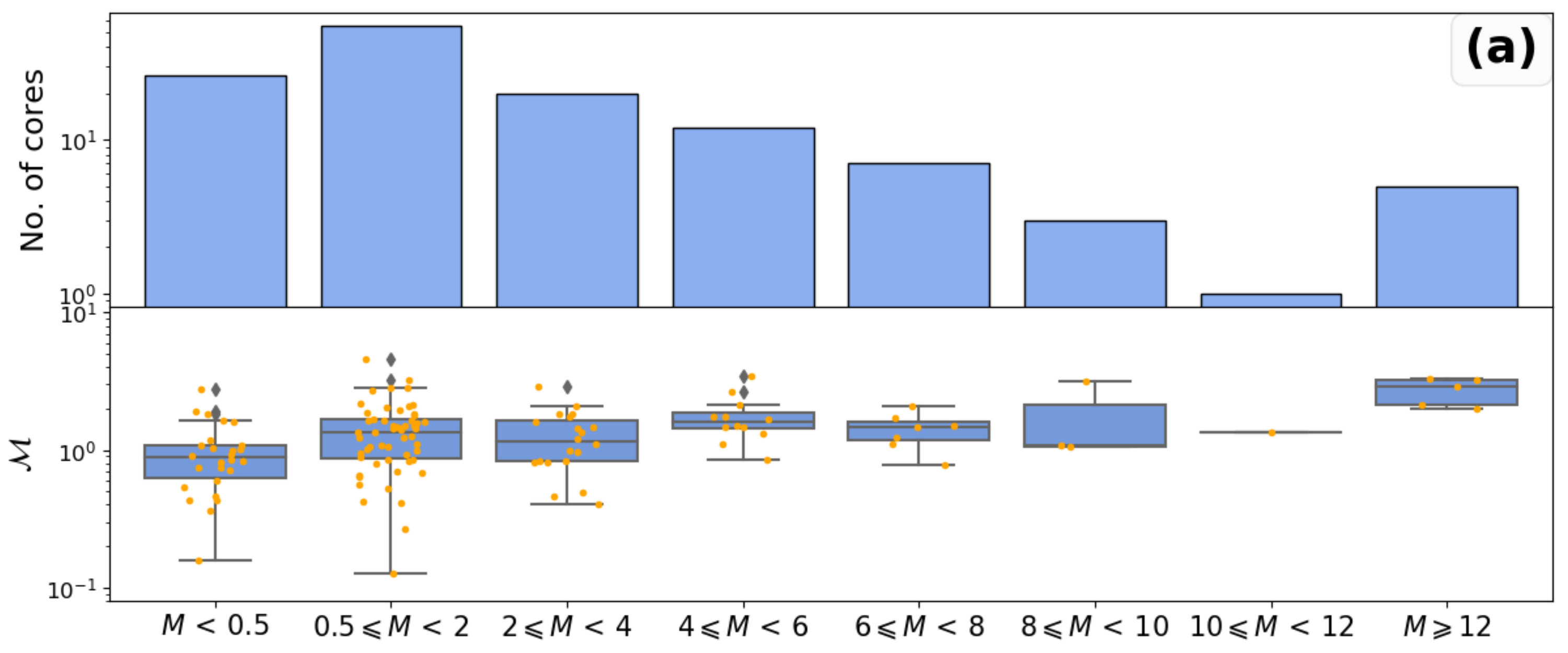}
\includegraphics[trim={0cm 0cm 0cm -1.5cm}, 
clip,width=0.95\textwidth]{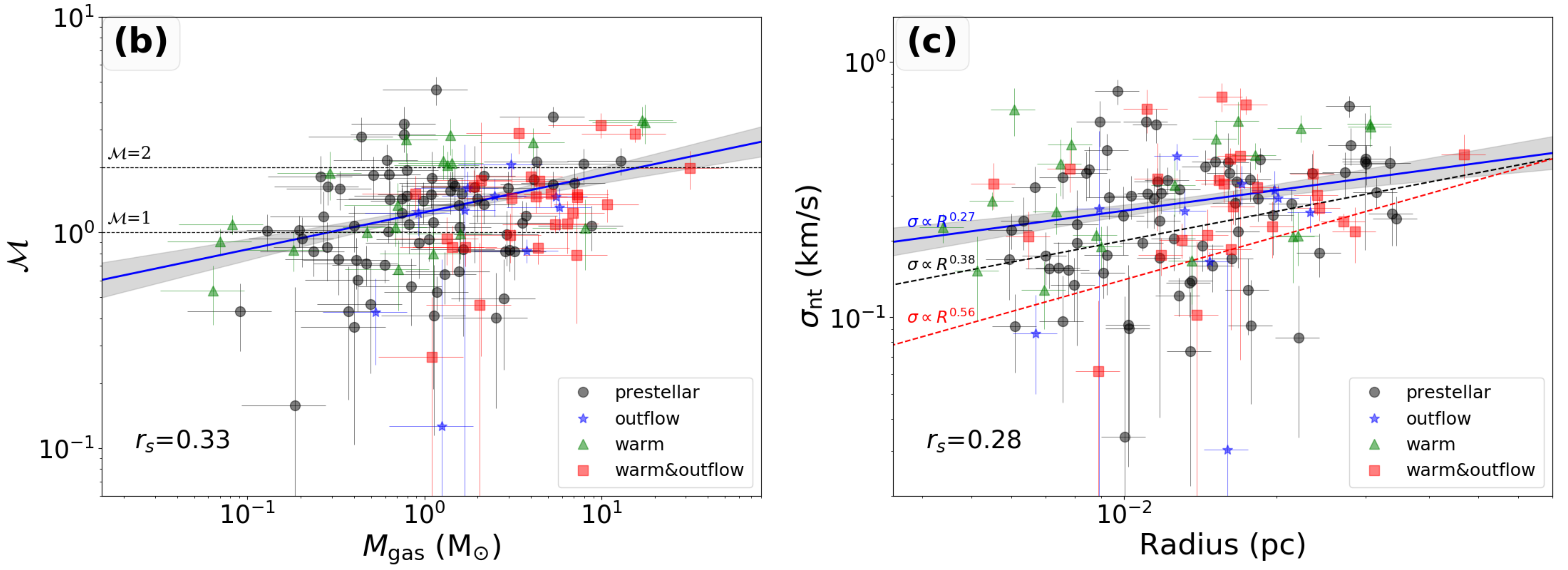}
\caption{
Panel (a): upper panel shows the 
number of sources per mass bin, and the bottom 
panel shows a box plot of the values of the Mach 
number per bin. In the box plot the mean value of the 
Mach number is shown with a horizontal line. 
Panel (b): Mach number versus mass for 
all the cores where molecular emission from \n2dp 
or \dcop was detected. 
The blue solid line shows the linear regression to all
the points in the plot (slope of 0.17 $\pm$ 0.04)  
and the gray shadowed area 
shows the 1$\sigma$ confidence interval for the fit.
The Spearman's rank test gives a coefficient of 
$r_{s}$ = 0.33, suggesting a moderate correlation 
between Mach number and gas mass.
Panel (c):  nonthermal velocity versus core radius 
for all the cores where molecular emission from \n2dp 
or \dcop was detected. The correlation between 
nonthermal velocity and core radius is weak, with 
a spearman's rank coefficient of $r_{s}$ = 0.28. 
The black dashed line shows the the original Larson 
relation with  $\sigma \propto R^{0.38}$ 
\citep{Larson-1981}, and the red  
dashed line is the revised \cite{Heyer-2004} relation 
with $\sigma \propto R^{0.56}$. 
The blue solid line indicates the best fitting result with 
$\sigma \propto R^{0.27}$. 
} \label{fig:mach}%
\end{figure*}
%%------------------------------------------------

\subsection{Turbulence}
\label{sec:turb}
\subsubsection{Global turbulence in IRDCs}
The origin of turbulence in molecular clouds is still 
not well understood, as well as its behavior with cloud 
evolution. Molecular cloud formation and destruction 
can be a dynamical process, in which  turbulence is 
transient, decaying quickly on timescales comparable 
to the cloud lifetime \citep[e.g.,][]{Elmegreen-2000,
Hartmann-2001,Dib-2007}. Alternatively, the turbulence 
can decay slowly in a  quasi-equilibrium process, 
in which turbulence is fed into 
the molecular cloud by, for example, protostellar 
outflows, \HII regions, or external cloud shearing  
\citep[e.g.,][]{Shu-1987,McKee-1999,Krumholz-2007,
Nakamura-2007}.

Simulations of isothermal molecular clouds with and 
without a continuous injection of energy have shown 
differences in the properties of the core gas internal 
velocity dispersion and the relative 
motions (i.e., the core-to-core centroid velocity dispersion) 
between the cores \citep[][]{Offner-2008,Offner-2008b}. 
Therefore, observations of these 
properties can in principle be used to analyze  the 
degree of turbulence within molecular clouds. 

In the driven scenario, i.e., where there is a continuous 
injection of energy into the cloud, the simulations show 
that, in a clump, prestellar cores have in average 
$\sigma_{\rm nt} < 1.5 c_{\rm s}$, and there is a small 
increment of this value for protostellar cores 
\citep{Offner-2008b,Offner-2008}. In the decay scenario, 
where there is no additional injection of energy to the 
cloud, prestellar cores have in average 
$\sigma_{\rm nt} < 1.0 c_{\rm s}$, and it increases 
to larger values for protostellar cores 
\citep{Offner-2008b,Offner-2008}. This is because, 
as the turbulence decays, the cloud will start to collapse,  
and material will fall into the clump gravitational potential.

In our observations, the mean velocity dispersion, 
derived from the \n2dp or \dcop, between the prestellar 
and protostellar cores increases slightly from 
$\sigma_{\rm nt} \sim (1.33 \pm 0.04) c_{\rm s}$ to 
$\sigma_{\rm nt} \sim (1.47 \pm 0.06)  c_{\rm s}$. 
This trend is also seen in 7 out of 9 clumps  
that have available $\sigma_{\rm nt}$ from \n2dp 
or \dcop.  G327.11 and G343.48 do not show this trend.    
There are three clumps (G331.37, G340.17, G340.22) 
with no available 
$\sigma_{\rm nt}$ from \n2dp or \dcop for the 
protostellar sample.   The velocity dispersion 
in the protostellar cores is similar or slightly 
smaller than those of prestellar cores in the 
latter two clumps. 
In addition, the mean velocity dispersion 
obtained from C$^{18}$O 
also shows a small increment from 
prestelllar to protostellar stages in a clump. 
The statistics for the  velocity dispersion of 
\dcop and \n2dp for each clump are presented in 
Appendix~\ref{app:A}  (see Figure~\ref{fig:hmapsig}). 
Overall, the measured core-averaged $\sigma_{\rm nt}$ 
does not increase significantly from prestellar to 
protostellar phases, suggesting that our observations 
are consistent with simulations where there is a 
continuous injection of energy, consistent with 
the driven scenario.

An additional piece of information can be obtained by 
studying the core-to-core velocity dispersion. 
For simulations where no additional sources of turbulence 
are injected, and therefore the turbulence is decaying, 
the relative motions between protostellar cores have a 
higher dispersion compared to the prestellar cores 
 \citep{Offner-2008b}. 
On the other hand, if the turbulence is driven, i.e., energy 
is continuously injected into the simulations, the prestellar 
cores have a higher core-to-core velocity dispersion than 
the protostellar cores \citep{Offner-2008b}.

The core-to-core velocity dispersion 
($\Delta v_{\rm LSR}$)  is computed from 
the standard deviation of the centroid velocity 
($v_{\rm LSR}$) of all of the dense 
cores within each clump. The calculations were performed 
 for clumps having more than two core centroid velocities 
available. We calculated the core-to-core centroid 
velocity dispersion between the cores for each clump
from the $v_{\rm LSR}$ of the Gaussian fits to the 
\dcop and \n2dp emission.  Figure~\ref{fig:hmap} 
shows a heat-map of the velocity dispersion for all 
the cores, and for the cores classified as prestellar 
and protostellar. We find that in three clumps 
the core-to-core velocity dispersion is  comparable 
between protostellar and prestellar cores 
(i.e., G28.27, G343.48, and G327.11), 
which is neither consistent with the driven nor 
the decaying turbulence simulations.  
The core-to-core velocity dispersion of prestellar 
cores is higher than that of protostellar cores in 
three clumps (i.e., G10.99, G332.96, and G340.23), 
which is consistent the driven turbulence scenario. 
On the other hand, we also find that the 
core-to-core velocity dispersion of prestellar 
cores is smaller than that of protostellar cores 
in other three clumps (i.e., G14.49, G337.54, 
and G341.03), which is consistent with the decaying  
turbulence scenario. 
The core-to-core velocity dispersion variation 
among the clumps might reflect differences in 
their internal turbulence, which possibly  
varies clump by clump and/or related to the 
local cloud environments. 
The statistics for the  core-to-core velocity dispersion 
derived from  
\dcop and \n2dp for each clump are presented in 
Appendix~\ref{app:A}  (see Figure~\ref{fig:hmapvlsr}).
Overall, the prestellar cores have a smaller 
core-to-core velocity dispersion compared to 
the protostellar cores if we take the mean 
value of all the ASHES clumps 
(Figure \ref{fig:hmap}).

The observations presented here, corresponding to 
the pilot survey \citep{Sanhueza-2019}, give some 
support to the scenario 
in which the global turbulence of IRDCs is driven,  
but a larger sample is required 
before to draw firm conclusions on the relative merits 
of driven or decaying turbulence. We expect to 
address this point again using the complete survey 
in the future.

%%--------------------------------------------
\begin{figure*}
\centering
\includegraphics[width=0.85\textwidth]{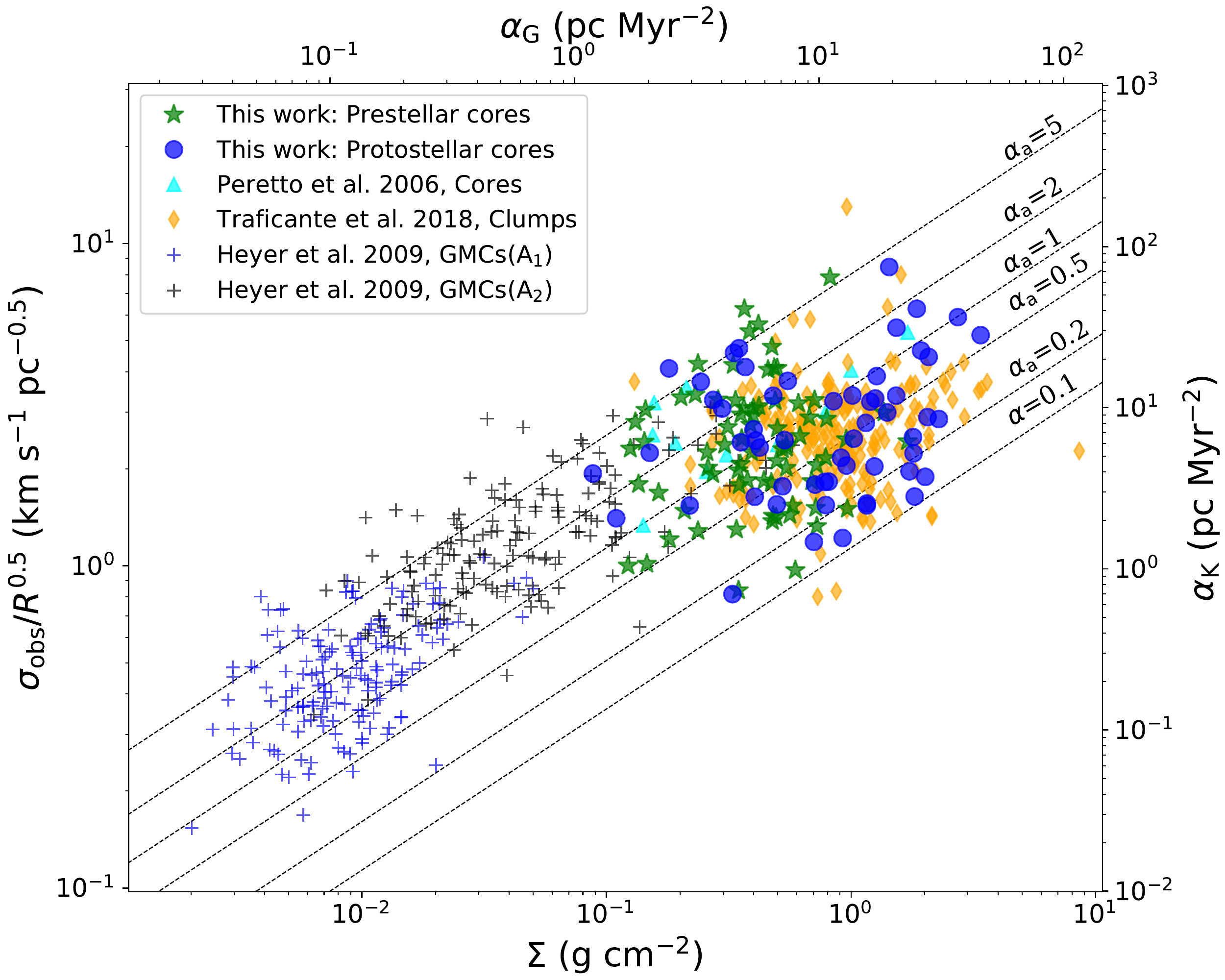}
\caption{Modified Larson relationship, also called 
Heyer relationship, between the cloud line widths, 
radius, and surface density. Top and right axes  
express the same quantities in term of acceleration 
parameters ($a_G,a_K$). Green stars and blue 
circles show the prestellar and protostellar cores, 
respectively. 
Cyan triangles present the values for the dense 
cores in high-mass star formation regions 
\citep{Peretto-2006}. 
Diamonds show the values for the massive 70 \um 
dark clumps \citep{Traficante-2018}. 
Crosses present the values obtained for the GMCs 
in the Galactic plane \citep{Heyer-2009}. 
}\label{gravoturbulence}
\end{figure*}
%%--------------------------------------------

\subsubsection{Turbulence within the cores}
To determine the turbulence of the gas within the cores, 
we analyzed the nonthermal velocity dispersion of the 
molecular emission detected from each core, and their 
Mach number (see Section~\ref{sec:mach}). 

The Mach number, which provides an indication of the 
level of gas turbulence, shows a moderate positive 
correlation with the mass of the cores (see panel (b) 
of Figure \ref{fig:mach}),  with a Spearman's rank 
correlation of $r_{s}$ = 0.33 
($p$–value=1$\times10^{-4}$).   
A similar positive correlation has also been 
seen in the IR-dark,  high-mass star-forming 
region NGC6334S \citep{Li-2020a}. 
Also, there is a small increment (11\%) in the mean 
Mach numbers from prestellar cores, with 
$\langle \mathcal{M} \rangle=1.33 \pm 0.04$,  
to protostellar cores, 
with $\langle \mathcal{M} \rangle=1.47 \pm 0.06$.

The nonthermal velocity dispersion derived from the 
deuterated species also slightly increases from the 
prestellar to the protostellar stage.  
The variation is small (14\%) from the prestellar cores, with 
($\langle \sigma_{\rm nt} \rangle =0.28$ \kms), to the protostellar 
cores, with ($\langle \sigma_{\rm nt} \rangle=0.32$ \kms). 
The nonthermal velocity dispersion of the envelope traced 
by C$^{18}$O remains essentially the same, with only a 
small variation of 12\%,  between the prestellar 
($\langle \sigma_{\rm nt,C^{18}O} \rangle=0.83$ \kms) and 
protostellar ($\langle \sigma_{\rm nt,C^{18}O} \rangle=0.93$ \kms) 
cores. 
This suggests that the variation of turbulence within 
the cores from the prestellar to the protostellar phase  
is negligible in the very early stages of high-mass star 
formation studied in ASHES, likely because not enough 
time has passed from the protostellar cores to inject 
turbulence into the surrounding medium (core and envelope).
This is consistent with the outflow analysis made on the 
ASHES sample by \cite{Li-2020b}; they presented a 
detailed study of the energy 
input given by the outflows and found that the outflow-induced 
turbulence contribution to the 
internal clump turbulence at the current epoch is very limited.

Studies from more evolved high-mass star-forming 
regions \citep[e.g.,][]{Sanchez-Monge-2013}, however, 
have shown larger differences between the prestellar 
and protostellar cores' line widths ($\sim$80\% 
differences); the mean observed velocity dispersion 
is 0.42  and 0.76~\kms for the prestellar and 
protostellar cores, respectively.  Larger values 
seen in more evolved objects seem to be 
associated with the effect of turbulence injected by 
outflow activity \citep[e.g.,][]{Liu2020}. 
This difference could be explained by 
the fact that the ASHES IRDCs are in a very early 
stage of evolution (70 \um dark), and what we define as 
protostellar cores in this sample corresponds to an 
intermediate stage, between prestellar cores and 
more evolved protostellar cores that already have 
significant emission at infrared wavelengths. 
This suggests that the outflow of these early 
protostellar cores is not yet significantly affecting 
the turbulence within the cores \citep[see also][]{Li-2020b}. 

The lack of a significant increment in the turbulence 
within the cores as they evolve suggests that the 
star formation activity in our protostellar cores is very 
recent (see also Figure~\ref{fig:hmap}). 
It is likely that the effect of the new, recently 
formed protostars have not had sufficient time and 
sufficient energy to considerably affect their natal 
cocoons at the core scale (a few 1000 au), and their 
effect may be localized at much smaller scales than 
those we can resolve with the current observations. 
The massive cores tend to have larger radii 
(see Figure~\ref{fig:MR}), 
whereas the nonthermal velocity dispersion  
has no significant change with the core radius 
(see Figure~\ref{fig:mach}). This suggests that the 
larger Mach number found in the massive cores is not 
because of the spectrum averaged over the larger 
core radius. 
Therefore, the increase in turbulence with the 
core mass might not be the result of core evolution, 
but rather a natural difference between low- 
and high-mass cores. 
The high-mass cores show relatively lower virial parameters, 
indicating that the high-mass cores would need additional 
support, even though it is not sufficient to 
maintain equilibrium, to counterbalance gravity, slowing 
down the collapse on times larger than the free-fall time.

\subsubsection{Are observed line widths dominated by gravitational collapse?}
\label{sec:linegravi}
If a cloud is undergoing global gravitational collapse, 
then the  line widths can be dominated by 
the inward motions produced by this collapse. This can 
happen at both cloud and/or core scales 
\citep{vazquez-2009,Ballesteros-2011}. 
Indeed, \citet{Heyer-2009} showed that the dynamics of 
molecular clouds can be dominated by the global 
gravitational collapse and presented a scaling relation 
that extends  the initial relation of \citet{Larson-1981}, 
between the clumps' line velocity dispersion and radius. 
The scaling relation proposed by \citet{Heyer-2009} 
shows that the ratio $\sigma/R^{0.5}$ is 
proportional to the surface density, $\Sigma$, of the 
clouds. 

The effect of the global collapse on the nonthermal 
motions can be estimated by a gravitational parameter 
that depends on the surface density ($\Sigma$) of the 
cloud and/or core, $a_{\rm G}=\pi G\Sigma/5$, compared to a 
kinetic parameter that depends on the line width 
($\sigma$) and radius (R) of the cloud and/or core 
$a_{\rm K}=\sigma_{\rm tot}^2/R$ 
\citep{Traficante-2018}. These two parameters are 
related to a virial parameter ($\alpha_a$) via 
$a_{\rm K}=\alpha_{\rm a} a_{\rm G}$, whose 
interpretation is different than the usual virial parameter 
derived from the ratio between virial mass and total 
gas mass (Section~\ref{sec:virial}). 
If the gravitational collapse dominates 
over the kinetic energy in the cloud, 
then $a_{\rm K}<a_{\rm G}$.

We computed these parameters for the clumps and their 
embedded cores from the ASHES survey, and compared  
them with the values found by \citet{Traficante-2018c} in a 
sample of clumps,  by \cite{Peretto-2006} for a sample of 
cores embedded in a high-mass star-forming region, and 
by \cite{Heyer-2009} for a sample of giant molecular 
cloud (GMC) in the Galactic plane  (see 
Figure~\ref{gravoturbulence}), which includes GMCs 
(A$_{1}$) and their high-density area within 1/2 
maximum isophote of H$_{2}$ column density (A$_{2}$).  

There are 47 out of 129 cores with a value of  
$\alpha_{a}$ smaller than 1 in the ASHES survey. 
In general, the cores have values of $a_{\rm G}$ and 
$a_{\rm K}$ leading to values of $\alpha_{\rm a}$ 
between 0.1 and 5.0, and  the dense clumps have a 
similar distribution.  On the other hand, the GMCs 
have relatively larger $\alpha_{\rm a}$ compared to the 
clumps and the cores.

To determine any real correlation, we 
calculated the Spearman's correlation 
coefficient 
between $\sigma_{\rm obs}/R^{0.5}$ 
and $\Sigma$ (or $a_{\rm G}$ and $a_{\rm K}$) for 
the prestellar and protostellar cores. 
The correlation coefficient for prestellar cores is 
$r_{s}$ = 0.05 ($p$–value=0.65), suggesting that 
there is no linear 
correlation between these values. For protostellar 
cores, the value of the correlation coefficient is 
$r_{s}$ = 0.27 ($p$–value=0.04), suggesting a moderate 
linear trend. This suggests that for prestellar 
cores the observed  line widths are not dominated 
by  gravity. However, once star formation begins, 
the gravitational collapse plays a role in 
the dynamics of the cores and thus contributes 
to the line widths. 
This is also consistent with the low value of the 
``standard" virial parameter found for the 
protostellar cores in which the role of 
gravitational potential energy with respects to the 
kinematic energy becomes more important  
(Section~\ref{stability-cores}).

\subsubsection{The first Larson relation: velocity dispersion-size relation}
\citet{Larson-1981} found an empirical power-law 
relationship between global velocity dispersion, 
$\sigma_{\rm obs}$ (\kms), and cloud size, $R$ (pc), 
of molecular clouds, with a slope of 0.38, 
$\sigma_{\rm obs} \propto R^{0.38}$. 
This correlation resembles the turbulent cascade 
of energy, and thus, it is considered as an indication 
that turbulence dissipates from large spatial scale 
down to smaller spatial scales.  
The $\sigma_{\rm obs}$--$R$ relation has also been found 
in the other studies of different molecular clouds, but 
with a different slope \citep[e.g., 0.56;][]{Heyer-2004}. 
On the other hand, the relation seems to break in 
some studies of star-forming regions, either with a 
significantly lower slope, e.g., 
$\sigma_{\rm obs} \propto R^{0.21}$
obtained from Orion A and B \citep{Caselli-1995}, 
or without correlation \citep{Plume-1997,
Ballesteros-2011,Traficante-2018}.

\begin{figure}
\centering
\includegraphics[width=0.45\textwidth]{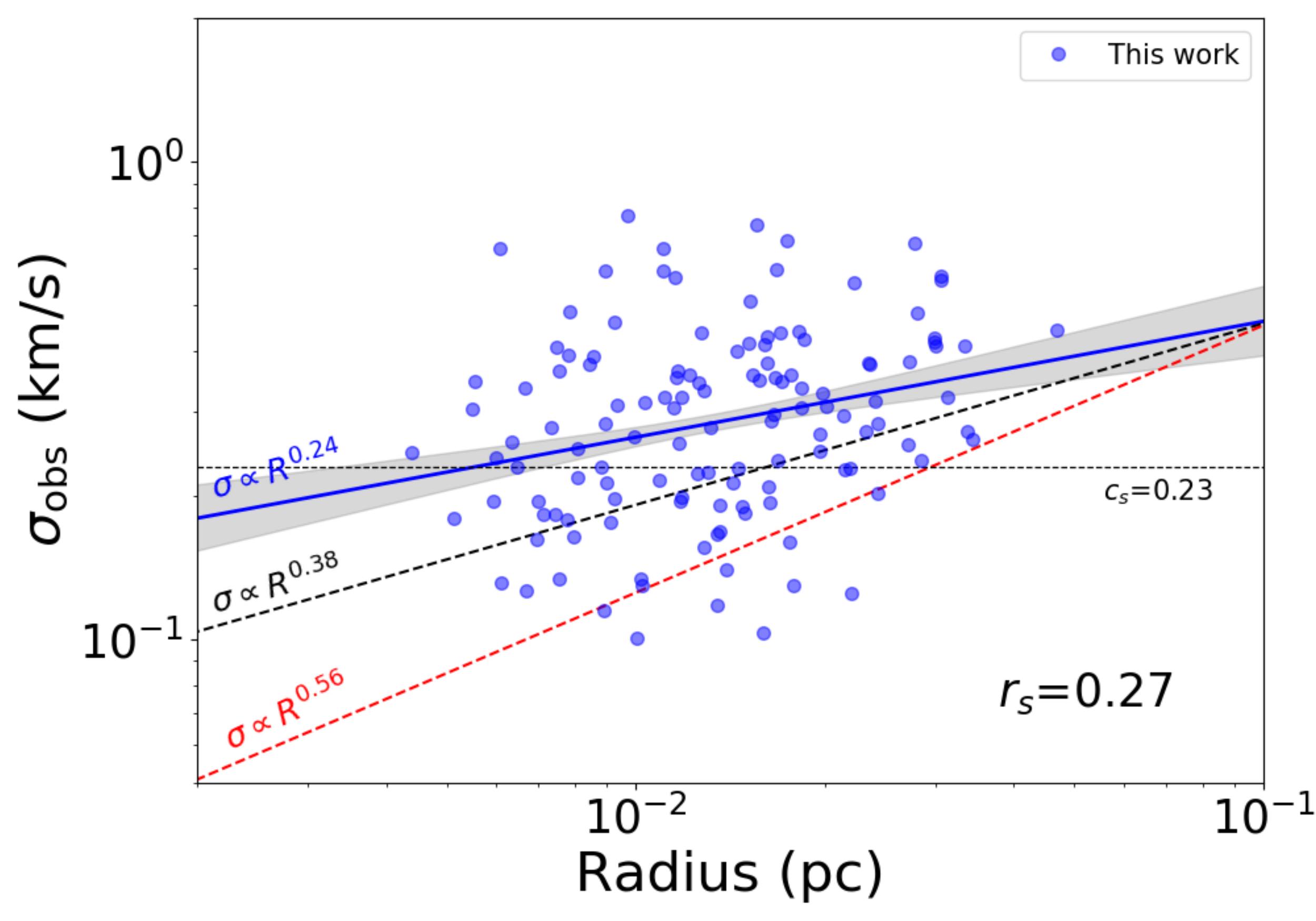}
\includegraphics[width=0.45\textwidth]{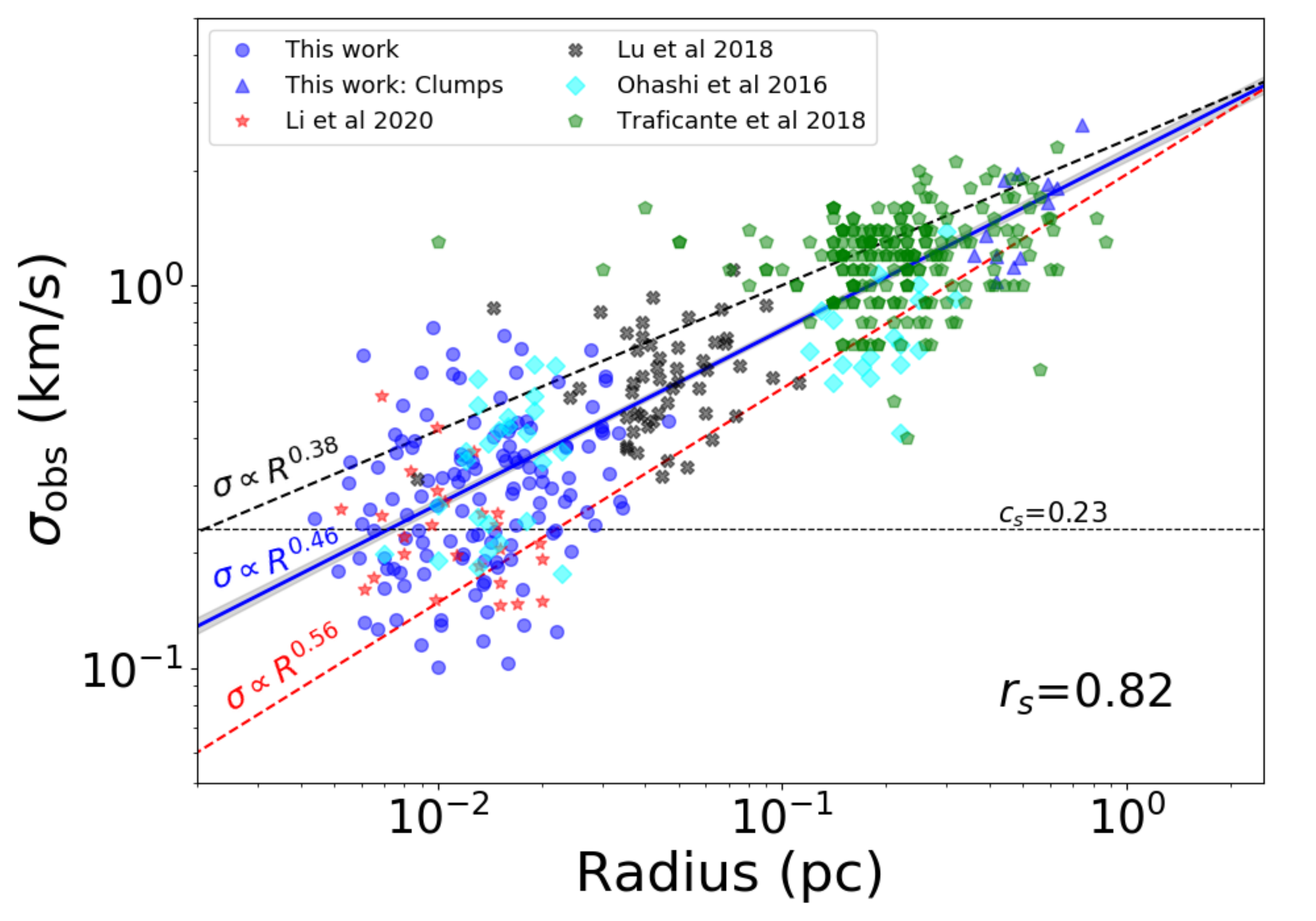}
\includegraphics[width=0.45\textwidth]{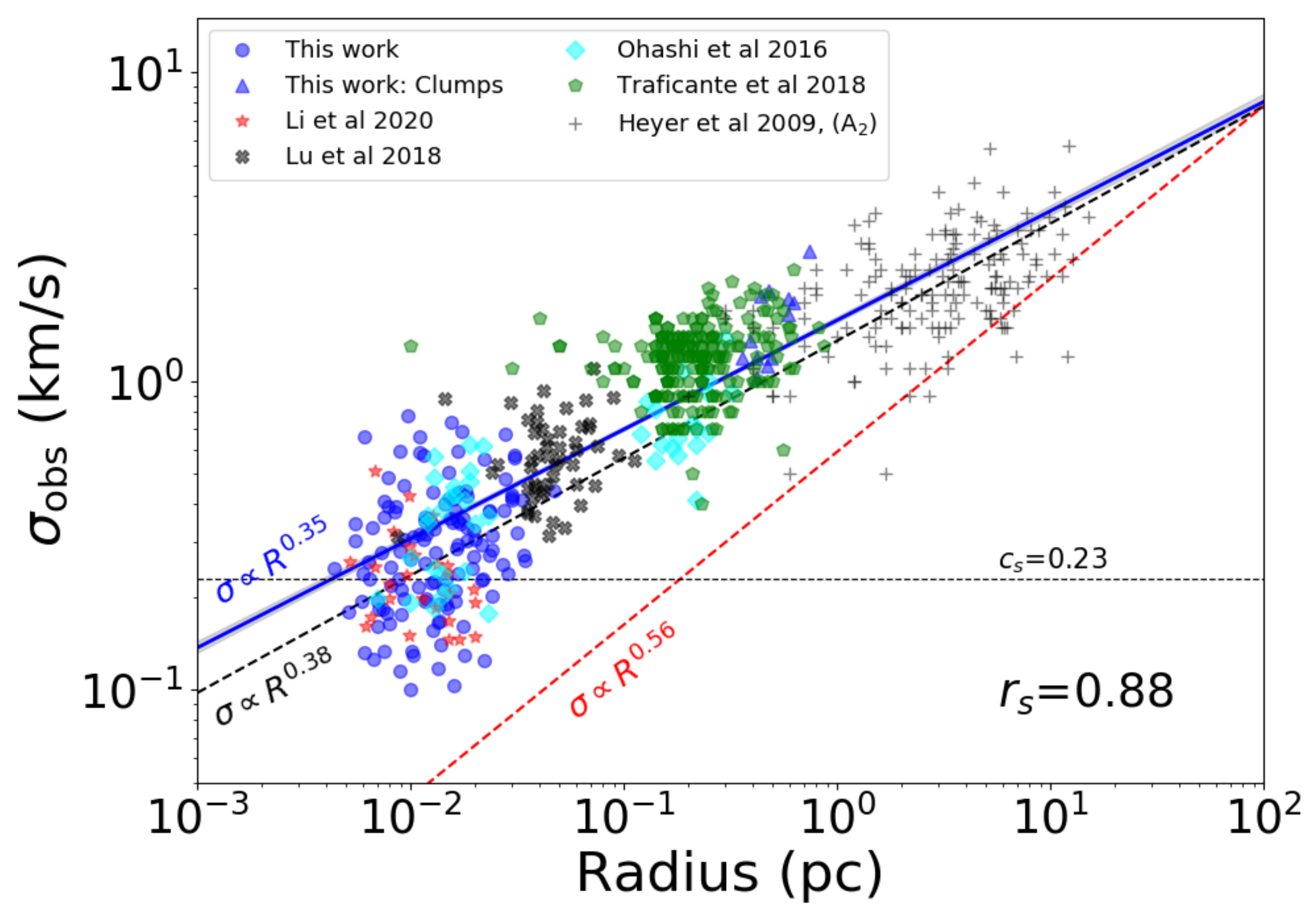}
\caption{Observed velocity dispersion $\sigma_{\rm obs}$ 
as a function of the radius. 
From upper to bottom, panels show the 
$\sigma_{\rm obs}$ versus radius for dense cores, 
clumps and dense core, and GMCs down to core, 
respectively.   
The blue solid line shows the linear regression to all 
the points in the plot,  and the gray shadowed 
area shows the 1$\sigma$ confidence interval for the fit. 
The best fit returns slopes of 0.24 $\pm$ 0.08, 
0.46 $\pm$ 0.01, and 0.35 $\pm$ 0.01 for upper, 
middle, and bottom panels, respectively. 
The Spearman's rank coefficients between observed 
velocity and radius are presented in the lower right 
corner in each panel. 
The black dashed line shows the the original Larson 
relation with  $\sigma_{\rm obs} \propto R^{0.38}$ 
\citep{Larson-1981}, and the red  
dashed line is the revised \cite{Heyer-2004} relation 
with $\sigma_{\rm obs} \propto R^{0.56}$. 
}\label{fig:sigrad}
\end{figure}

To explore the variation of $\sigma_{\rm obs}$--$R$ 
relation in different 
spatial scales, we have retrieved data from \cite{Li-2020a}  
and \cite{Lu-2018} for a sample of embedded cores in 
high-mass star-forming regions, from \cite{Ohashi-2016} 
for a sample of high-mass clumps and embedded cores, 
from \cite{Traficante-2018c} for a sample of 
high-mass clumps, 
and from \cite{Heyer-2009} for a sample of GMCs. 
From the top panel of Figure~\ref{fig:sigrad}, we note that 
there is a weak correlation ($r_{s}$ = 0.27,  
and $p$–value=0.002) between 
$\sigma$ and R for ASHES cores only, with a slope of 
0.24 $\pm$ 0.08 that is similar to the values of 0.2--0.3 
reported in \cite{Caselli-1995} and \cite{Shirley-2003}. 
For ASHES cores only, the sample covers a limited 
range of radii, which could break down the 
$\sigma_{\rm obs}$--$R$ 
relation.

As shown in the middle panel of Figure~\ref{fig:sigrad}, 
if the clumps and relatively larger size cores are included, 
the $\sigma_{\rm obs}$ and $R$ show a strong correlation 
($r_{s}$ = 0.82,  and $p$–value=4.5$\times10^{-114}$), 
with a slope of 0.46 $\pm$ 0.01 that 
is between 0.38 and 0.56.  
On the other hand, if the GMCs are included, the 
correlation between $\sigma_{\rm obs}$ and R becomes 
even stronger ($r_{s}$ = 0.88, and 
$p$–value=2.7$\times10^{-208}$), but  
however, the slope 
(0.35 $\pm$ 0.01)  becomes more flattened. 
The change in the slope between including or excluding  
GMCs could be due to intrinsic turbulent variation 
from GMC scale to clump and/or core scale or observational 
bias since the measurements are obtained with  different 
molecular lines and different kind of telescopes. 
The GMCs values were derived from $^{13}$CO, while 
clumps and/or cores were obtained from more dense gas 
tracers (i.e., NH$_{3}$, \dcop, and \n2dp).  
Unfortunately, we can not conclude which effect  
drives the variation in the slope. 

Overall, our results suggest that the relation between 
$\sigma_{\rm obs}$ and $R$ persists from GMC scale down to 
core scale (Figure~\ref{fig:sigrad}), and the slope is 
between  1/3 and 1/2, the former is expected for 
turbulence-dominated solenoidal motions, and the 
latter expected for shock-dominated turbulence 
\citep{Kolmogorov-1941,Galtier-2011,Federrath-2013}. 
Our results also indicate that caution is required in 
interpreting the investigation of a sample with a 
small range of radii because it could break down 
the $\sigma_{\rm obs}$--$R$ relation due to the small dynamical 
range in the spatial size.

%%--------------------------------------------
%%--------------------------------------------

\subsection{The relationship between the embedded cores 
and parental clump}
\subsubsection{Influence of the parental clump properties on the cores properties} 
%\label{sec:corekine}
%
We analyze whether the global properties of the 
clumps, such as their mass, virial mass, surface density, 
velocity dispersion, and virial parameter, have any impact 
on the properties of their embedded cores. 
To quantify any correlation, we calculated the Spearman's  
correlation coefficients between different parameters 
of the clumps and the cores.

We found a strong correlation between the mass 
of the clump and the fraction of protostellar cores 
($r_{s}=0.50$, and $p$–value=0.095). There is also a strong    
anticorrelation between the surface density of the 
clumps and the average value of the core's virial 
parameter ($r_{s}=-0.76$, and $p$–value=0.004), 
suggesting that, at a higher 
surface density, clumps encompass more 
gravitationally bound cores within it. 
There is no correlation between the number of 
embedded cores and the clump velocity dispersion 
($r_{s}=-0.05$, and $p$–value=0.89), 
indicating no significant relationship 
between the turbulence 
within the clumps and their level of fragmentation. 
We found a moderate correlation between the global virial 
parameter of the clump and the mean value of the 
virial parameter of the cores ($r_{s}=0.36$, 
and $p$–value=0.25). 
We also found a weak correlation ($r_{s}=-0.29$, 
and $p$–value=0.35) 
between the virial parameter of the clump and the 
fraction of cores per clump that are gravitationally 
bound ($\alpha<2$), and a weak correlation 
($r_{p}=-0.21$ and $p$–value=0.51) 
is seen between the virial 
parameter of the clump and the number of 
embedded cores. 

We also analyzed whether the velocity dispersion of the 
clumps increased with the number of 
embedded protostellar cores, as one may expect if the 
turbulence is injected from the protostellar objects into 
the surrounding medium. However, we find a very weak  
correlation between the number of protostellar cores 
and $\sigma_{\rm clump}$ ($r_{s}=-0.11$, and 
$p$–value=0.74). Therefore, 
the turbulence injected by the protostellar cores in 
form of outflows is not enough to increase the overall 
turbulence in the clumps at these early stages of 
evolutions. This is consistent with the results of 
\cite{Li-2020b}.

Overall, we found no strong correlation between the 
large-scale properties of the clumps with the 
small-scale properties of the cores, except for the mass 
and surface density of the clump that may have an  
impact on the properties of their embedded cores.

\subsubsection{The kinematical relationship between cores and the parental clump} 
\label{sec:corekine}
% %
On average, the core-to-core velocity dispersion 
($\Delta v_{\rm lsr}$ = 0.5--2.9 \kms) derived from the 
\co18 line shows the highest value among all detected 
lines, except for CO. 
The core-to-core velocity dispersion  
is 0.4--1.8 \kms for \h2co and 0.2--1.6 \kms for \ch3oh, 
both higher than those of  the  \n2dp and \dcop lines; 
0.2--1.1 \kms for \n2dp and 0.2--1.0 \kms for \dcop.
The latter two lines  have comparable $\Delta v_{\rm lsr}$ 
in most of the clumps.   
The discrepancy of  $\Delta v_{\rm lsr}$ in different lines  
is most likely because they trace different gas 
components \citep[e.g., \h2co, \ch3oh, 
CO and its isotopologues  are easily affected by 
protostellar outflows,][]{Li-2020b,Li-2022b} 
The core-to-core velocity dispersions derived from \n2dp 
and \dcop are comparable to that found in low-mass 
star-forming regions, such as the core-to-core 
N$_{2}$H$^{+}$ velocity dispersion for Perseus 
\citep[$\sim$0.4~\kms;][]{Kirk-2010}.

The comparison between core-to-core motions and the 
global motions of the clumps can be used to evaluate 
how the dense cores are connected to the lower-density 
gas in the clumps \citep[e.g.,][]{Kirk-2010}. 
For example, if the dense cores are connected to the 
large-scale motions, then the dispersion of the core centroid 
velocities should be similar to the global velocity dispersion 
of the natal clump. On the other hand, we expect to see a 
much smaller core-to-core velocity dispersion than the 
global velocity dispersion of the natal clump if  the dense 
cores are kinematically detached from the large-scale 
motions within the clump.  
The core-to-core velocity dispersions obtained from 
the \n2dp and \dcop lines are about 4 times smaller than 
the derived \co18 velocity dispersion 
($\sigma_{\rm obs,C^{18}O}$) over the clump 
that is also about 2--3 times higher than the core-to-core 
velocity dispersions derived from \h2co and \ch3oh. 
These results suggest that the dense cores are 
kinematically detached from the large-scale motions 
(e.g., large-scale gas flows).

\subsection{The relative motions of dense cores and their envelopes} 
\label{sec:relmotion}
Simulations of star formation suggest that the mass of a 
star could be determined by the motions of its path 
through the natal molecular cloud \citep{Bonnell2001}. 
The relative motions between dense cores and their associated 
envelopes can be examined using suitable molecular 
lines, e.g., a high-density gas tracer can be used to probe 
the dense core, and a low-density gas tracer can be utilized 
to measure the envelope. 
In this work, we used \n2dp and \dcop (hereafter 
deuterated species) as our high-density 
core tracers, because both molecules have high critical 
densities ($1.7-1.8 \times 10^{6} \, \rm cm^{-3}$), 
and used \co18 as the low-density envelope tracer 
because it has a low critical density 
($9.3 \times 10^{3} \, \rm cm^{-3}$).  
The line-center velocity ($v_{\rm LSR}$) differences 
between deuterated 
species and \co18 range from 0.24  to 1.85 \kms 
(see also Tabel~\ref{tab:vlsr}),  
with mean and median values of 0.64  and 
0.41 \kms, respectively.  
Our observed values are greater than that of 
$\sim$0.1 \kms measured toward low-mass 
star-forming regions \citep{Walsh2004}.  
The derived differences in line-center velocities are 
smaller than the deuterated species and \co18 
line widths (FWHM; see Table~\ref{tab:vlsr}), 
which is similar to the results of 
low-mass star-forming regions \citep[e.g.,][]{Walsh2004}. 
The mean and median values of line-center velocity 
differences for each clump and the line widths of 
deuterated species and \co18 are tabulated in Table~\ref{tab:vlsr}.  
The line-center velocity differences are expected to be comparable  
to or even somewhat greater than the \co18 line width 
if cores move ballistically from their birth sites 
\citep[][and reference therein]{Walsh2004}. 
The observed small difference indicates that the dense 
cores are comoving with their envelopes rather than 
moving ballistically.  
In the framework of \cite{Bonnell2001}, this result 
might imply that the embedded 
dense cores do not obtain large amounts of mass by 
accreting material as they move through the lower-density 
environment in their natal molecular cloud.

\begin{table}
\scriptsize
%\tiny
\centering
\caption{Line-center velocity difference and line width}
\label{tab:vlsr}
\begin{tabular}{l c c c c c c c c c  c }
\toprule
Clump & 
\multicolumn{2}{c}{$v_{\rm LSR,deu}$ - $v_{\rm LSR,C^{18}O}$} & 
\multicolumn{2}{c}{FWHM$_{\rm deu}$} &
\multicolumn{2}{c}{FWHM$_{\rm C^{18}O}$} \\ 
\cmidrule(l{1pt}r{2pt}){2-3} \cmidrule(l{2pt}r{2pt}){4-5} 
\cmidrule(l{1pt}r{2pt}){6-7}
& median & mean 
& median & mean 
& median & mean  \\   
	\midrule    
G10.99 & 0.30  & 0.41  & 0.67  & 0.71  & 1.78  & 1.95 	\\ \hline 
G14.49 & 0.70  & 1.19  & 0.63  & 0.80  & 2.43  & 2.65 	\\ \hline 
G28.23 & 0.20  & 0.37  & 1.01  & 1.09  & 2.61  & 3.29 	\\ \hline
G327.11 & 1.54  & 1.85  & 0.68  & 0.68  & 1.81  & 2.42 	\\ \hline 
G331.37 & 0.17  & 0.30  & 0.70  & 0.74  & 1.43  & 1.60 	\\ \hline
G332.96 & 0.37  & 0.41  & 0.96  & 1.02  & 1.60  & 1.73 	\\ \hline
G337.54 & 0.72  & 0.99  & 0.48  & 0.64  & 1.78  & 2.17 	\\ \hline
G340.17 & 0.73  & 0.73  & 0.99  & 0.99  & 2.15  & 2.63 	\\ \hline
G340.22 & 0.24  & 0.30  & 0.75  & 0.79  & 2.03  & 2.53 	\\ \hline
G340.23 & 0.27  & 0.41  & 0.87  & 0.94  & 2.26  & 2.48 	\\ \hline
G341.03 & 0.28  & 0.46  & 0.58  & 0.63  & 1.91  & 1.87 	\\ \hline
G343.48 & 0.09  & 0.24  & 0.47  & 0.57  & 1.45  & 1.57 	\\ \hline
minimum & 0.09  & 0.24  & 0.47  & 0.57  & 1.43  & 1.57 	\\ \hline
maximum & 1.54  & 1.85  & 1.01  & 1.09  & 2.61  & 3.29 	\\ \hline
median & 0.29  & 0.41  & 0.69  & 0.77  & 1.86  & 2.29 	\\ \hline
mean & 0.47  & 0.64  & 0.73  & 0.80  & 1.94  & 2.24 	\\ \hline 
\bottomrule
\end{tabular}
\tablenotetext{}{Notes. 
The unit is \kms. 
}
\end{table}

\subsection{Properties of the most massive cores}
We found 6 cores that are relatively massive 
(10.80--31.66 $M_\odot$). On average, these cores have a 
core nonthermal velocity dispersion 1.7 times larger 
than the remaining lower mass cores (0.49 \kms 
compared to 0.29 \kms). 
On the other hand, the most massive cores have  
envelopes, traced by \co18, with a nonthermal 
velocity dispersion 1.4 times larger than the values of the 
lower mass cores (1.22 \kms compared to 0.86 \kms).  
The most massive cores are strongly self-gravitating 
($\langle \alpha \rangle=0.37$) with a higher surface 
density ($\langle \Sigma \rangle = 1.50$ g cm$^{-2}$) 
and higher peak column density 
($\langle N_{\rm peak}(\rm H_{2}) \rangle = 2.02 \times 10^{23}$ cm$^{-2}$), 
compared to 
the lower mass cores ($\langle \alpha \rangle=1.44$,  
$\langle \Sigma \rangle = 0.49$ g cm$^{-2}$, and 
$\langle N_{\rm peak}(\rm H_{2}) \rangle = 5.38 \times 10^{22}$ cm$^{-2}$). 
The most massive cores also have a mean Mach number of 
$\langle \mathcal{M} \rangle =2.5$ which is larger than those 
of lower mass cores of $\langle \mathcal{M} \rangle=1.3$.

The derived properties for the most massive cores in 
the ASHES sample are not only different from the lower 
mass cores in the same sample but also  
different from the cores embedded in low-mass 
star-forming regions, in terms of mass, surface density, 
and velocity 
dispersion \citep[e.g., Ophiuchus, Serpens, or 
Perseus;][]{Myers-1983,Caselli-2002,Kirk-2006,
Enoch-2008,Friesen-2009,Bovino-2021}. 
For low-mass star-forming regions, the 
typical core masses range between 0.1 and 
10 $M_\sun$ with a typical surface density 
of $<$ 1.0 g cm$^{-2}$ 
\citep[e.g.][]{Hogerheijde-1999, Kirk-2006,Enoch-2008}. 
In addition, the core nonthermal velocity dispersion is 
larger in our cores compared to those cores (typical size 
$\sim$0.06 pc) seen in low-mass star-forming regions 
\citep[typical nonthermal velocity dispersion of 
$\lesssim$0.2 \kms, e.g.][]{Caselli-2002,Friesen-2009,
Chen-2019}. 
We conclude that the physical properties of the 
most massive cores in IRDCs are inconsistent with 
cores that will only form low-mass protostars. 
Based on the evidence presented in this work, we 
suggest that, at least, the most massive cores in 
ASHES are likely to be the seeds of future high-mass stars.

\subsection{Comparison with high-mass star formation models}
\label{sec:msf}
Two high-mass star formation scenarios, 
``core-fed" and ``clump-fed", have been proposed 
in previous theoretical studies 
(see Section~\ref{sec:intro}). 
In this section, we compare our observational  
results with four representative high-mass star 
formation models in terms of the core's dynamics 
or properties.

In the ``core-fed" model, the mass of the star is 
determined by the mass reservoir of its parental 
core, which gathers the mass in a prestellar stage 
\citep{mckee-tan-2002}. 
Therefore, the high-mass prestellar core ($\gtrsim$ 30 \Mo),  
as the cornerstone of this model \citep[is known as 
``turbulent core accretion model,"][]{mckee-tan-2002}, 
must exist in order to form a  
high-mass star. 
However, based on our results, we do not find the high-mass 
prestellar core over 197 prestellar cores toward 12 
massive dense clumps. 
The most massive prestellar core  detected has a mass 
of  12.89 \Mo. 
In addition, we find that the majority of identified cores 
have virial parameters below the critical value of 2 for 
a nonmagnetized cloud, which is inconsistent with the 
quasi-equilibrium configuration proposed by the 
``core-fed" model \citep{mckee-tan-2002}. 
This indicates that the ``core-fed" model cannot explain 
the virial parameters of dense cores seen in our 
observations. We note that the limited amount of 
studies including estimations of the magnetic field 
strength in the virial equilibrium analysis indicate 
that the most massive cores still remain (strongly) 
subvirialized \citep[e.g.,][]{Beuther-2018,
JunhaoLiu-2020,Morii-2021,Sanhueza-2021}.

In the ``clump-fed" scenario, the final mass of stars is not 
determined by the mass of prestellar cores. In this  
scenario, the material can be replenished by funnelling 
mass from large scales (e.g., clump, cloud) to small 
scales (e.g., dense core). For instance, filaments can 
continue to provide the mass reservoir for the growth of 
stars.  Therefore, protostars embedded 
in cores can continue to grow in mass through accreting 
material from their parental clump and/or cloud.  
There are three representative ``clump-fed"  theoretical 
models, i.e., competitive accretion scenario  
\citep{Bonnell2001,bonnell-2004}, 
global hierarchical collapse 
\citep[GHC;][]{2017MNRAS.467.1313V,2019MNRAS.490.3061V}, 
and inertial-inflow model \citep{Padoan-2020},  
in which protostars would then be fed from the 
surrounding clump and/or cloud, in order to accumulate 
the mass to form high-mass stars.

In the competitive accretion model, fragmentation 
produces low-mass stars while the stars near the 
center of the gravitational potential well grow 
from low to high mass via accretion 
\citep{Bonnell2001,bonnell-2004}. 
To form a high-mass star from a low-mass stellar 
``seed"  in this model,  the embedded object maintains 
approximately a virial parameter of $\alpha$ = 0.5  
\citep{bonnell-2006}. 
As shown in Figure~\ref{fig:alpha}, most of the detected 
cores have virial parameters not around 0.5, 
which appears to be inconsistent with simulations 
of competitive accretion.

On the other hand, we also find some agreements  
between the simulation and the observations. 
We find that the subsonic turbulent motions of  
cores in competitive accretion simulations are in agreement 
with the measurements in identified cores where 
turbulence is dominated by subsonic or transonic motions  
\citep[see Section~\ref{sec:mach},][]{Padoan2001}.  
In addition, the line-center velocity difference between 
high-density cores is smaller than the \n2dp and 
\dcop line widths, which traces the high-density cores, 
and certainly smaller than the \co18 line width 
(see Section~\ref{sec:relmotion} and 
Table~\ref{tab:vlsr}), which 
traces the low-density envelopes.  
The small velocity difference is inconsistent with the 
competitive accretion model, in which dense cores 
have large velocities relative to the low-density 
envelopes due to the cores' movement within the 
clump \citep{Ayliffe2007}.

In the GHC scenario, the dense 
cores form from hierarchical gravitational fragmentation, 
in which mass is accreted from parent structures onto  
their embedded substructures 
\citep{2017MNRAS.467.1313V,2019MNRAS.490.3061V}.  
This model can be considered as an extension  of the 
competitive accretion model.  
The GHC model suggests that the nonthermal motions of the 
molecular clouds are dominated by infall and by a moderately 
supersonic turbulent background, which has a typical sonic 
Mach number of $\sim$3. 
As discussed in Section~\ref{sec:linegravi}, the gravitational 
collapse indeed plays a role in the dynamics of the protostellar 
cores, but not for prestellar cores. 
The 1D Mach number derived from \co18 is around 3--4.  
This is similar to the values of the GHC model, 
although the Mach number could be  
larger for the lower-density gas traced by CO; CO 
($n_{\rm crit} \sim 2 \times 10^{3} \, \rm cm^{-3}$) has 
a relatively lower critical density than that of \co18 
($n_{\rm crit}=9.3\times10^{3}$ cm$^{-3}$). 
In addition, the inverse $\alpha \, \sim \, M_{\rm gas}$ 
relation seen in our data can be explained by the GHC 
model, if the sample satisfies $M \, \propto \, R^{p}$ 
with $0 \, < \, p \, < \,3$; the detected cores show 
$M \, \propto \, R^{1.89}$ (see Figure~\ref{fig:MR}).
The observations are consistent with the GHC model in 
this respect.

In the inertial flow model, the dense cores are originated 
from turbulent fragmentation in supersonic turbulence  
environments, in which high-mass stars are assembled 
by large-scale, converging, and inertial flows 
\citep[][]{Padoan-2020}.  The infall rate of dense cores 
is controlled by the large-scale inertial inflow rather than 
the current core mass and density. 
In general, the accretion time increases 
with mass, indicating that more massive stars require 
longer accretion times than the less massive ones in 
order to achieve their final mass. This is in 
agreement with the observed outflow properties in 
the ASHES sample.  We find that the estimated outflow dynamical timescale 
increases with the core mass, which indicates that more 
massive cores have longer accretion timescales than 
less massive cores \citep[see][]{Li-2019b,Li-2020b}.  
In addition, the derived mass distribution of the prestellar 
core candidates is comparable to the prediction of mass 
distribution from the inertial flow model 
\citep{Padoan-2020,Sanhueza-2019,Morii-2023}.

In general, ASHES observations are more consistent with the 
``clump-fed" scenario rather than with the ``core-fed" scenario, 
although all ``clump-fed" models have their own 
limitations and cannot fully explain the dynamics of 
dense cores found in the observations. 
A more complete and realistic theoretical model of 
high-mass star formation should be able to reproduce 
(1) the observed dynamics of dense cores revealed in 
 ASHES, including the subsonic and transonic 
turbulence in dense cores, nonequilibrium state of 
dense cores, comoving of high-density cores and 
their low-density envelopes, and 
kinematically detached dense cores from large-scale motions;
(2) the characteristics of dense cores found in ASHES 
\citep[see][and references therein]{Sanhueza-2019}, 
including the presence of a large population of low-mass 
cores but without  high-mass prestellar cores, 
hierarchical subclustering, 
absence of primordial mass segregation, 
and a slightly shallower core mass function (CMF) 
than the slope of the Salpeter’s initial mass function (IMF).

\section{Conclusions}
\label{sec:con}
We analyzed the kinematics and dynamics of 12 IRDCs 
and their \ndustcores~embedded cores observed as part of the pilot ALMA Survey of 70 $\mu$m dark High-mass clumps in Early Stages (ASHES). For these, 
we determined their properties, including the velocity 
dispersion, gas mass, virial mass, virial parameter, 
Mach number, and core-to-core velocity dispersion. 
Our main results can be summarized as follows.

 1. The prestellar and protostellar cores have similar gas 
 temperatures obtained from NH$_{3}$, suggesting that 
 the protostellar cores in our sample are still at a very early 
 evolutionary phase. On the other hand, the gas mass, 
 column density, and volume density of the protostellar 
 cores are higher than those of the prestellar cores, 
 likely indicating a core mass growth from the prestellar 
 to the protostellar stage.    

2. On average, protostellar cores have relatively higher 
virial masses compared to prestellar cores. 
The virial parameter, $\alpha$, decreases with the mass 
of the cores. Therefore, the more massive cores are 
more unstable against gravitational collapse if one 
neglects the effects of magnetic field support. 
The virial parameter tends to decrease with the evolution 
of the cores. The value of $\alpha$ is slightly 
lower for protostellar cores, compared to the 
prestellar cores, suggesting they might be prone to 
gravitational instabilities. 
In addition, the most massive cores ($>$ 10 \Mo) are 
strongly self-gravitating ($\langle \alpha \rangle=0.37$), 
have higher nonthermal velocity dispersion 
($\langle \sigma_{\rm nt} \rangle=0.49$ \kms), 
higher surface 
density ($\langle \Sigma \rangle = 1.50$ g cm$^{-2}$), 
higher peak column density 
($\langle N_{\rm peak}(\rm H_{2}) \rangle = 2.02 \times 10^{23}$ cm$^{-2}$), and higher Mach numbers 
($\langle \mathcal{M} \rangle=2.5$) 
compared to the lower mass cores 
($\langle \alpha \rangle=1.44$, 
$\langle \sigma_{\rm nt} \rangle=0.29$ \kms,  
$\langle \Sigma \rangle = 0.49$ g cm$^{-2}$, 
$\langle N_{\rm peak}(\rm H_{2}) \rangle = 5.38 \times 10^{22}$ cm$^{-2}$, and $\langle \mathcal{M} \rangle=1.3$). 
The most massive cores in ASHES likely 
harbor the seeds of future high-mass stars.

3. There is a positive correlation between the Mach 
number and the mass of the cores. This suggests that 
the turbulence within the cores increases with their mass. 
In addition, most of the deuterated detected cores (84\%) 
are subsonic or 
transonic, with Mach number values between $0.1$ and 2.0. 
The level of turbulence within the cores, traced by their 
velocity dispersion and turbulent Mach number, does not 
increase significantly with the evolutionary stage of the core. 
This suggests that, once star formation begins, the core 
remains oblivious to the turbulence generated by outflows 
at the current extremely early evolutionary stage traced in 
these 70 $\mu$m dark IRDCs. 

4. With respect to the core-to-core velocity dispersion, 
and the core velocity dispersion, we found that 
the later is consistent with what is predicted in 
simulations 
of molecular clouds in which turbulence is continuously 
injected into the system, while the former is 
in disagreement with the driven turbulence and decaying 
turbulence scenarios.  

5. We also analyzed whether the line width in the IRDCs 
and cores can be dominated by the gravitational collapse. 
We find that the protostellar cores show a moderate 
correlation between $\sigma_{\rm obs}/R^{0.5}$ and 
$\Sigma$, indicating that the gravitational collapse 
plays a role in the turbulence of the protostellar 
cores and thus contributes to the line width. 
However, for the prestellar cores, there is no clear trend,  
which suggests that their line width is not dominated 
by gravity-induced motions.

6. We also find that the relation between velocity dispersion 
and size, also known as the first Larson relation, persists 
from the GMC scale down to the core scale. We note that considering 
a sample with a small range of radii can easily break 
down this relation.

7. Within each clump, the core-to-core velocity dispersion   
obtained from dense gas tracers is about 2--4 times 
smaller than the \co18 velocity dispersion over the entire 
clump.  This indicates that the dense cores are kinematically 
detached from the large-scale motions. 
In addition, the differences of line-center velocities 
in low (\co18) and high (\n2dp and \dcop) density 
tracers are smaller than the line width of \n2dp, 
\dcop, and \co18 lines. 
This indicates that the dense cores have small velocity shifts 
relative to their low-density envelopes, suggesting that 
the dense cores are comoving with their envelopes rather 
than  moving ballistically.

8. We compare our observational results to existing 
high-mass star formation models, including ``core-fed" 
(i.e., turbulent core accretion model)  and  ``clump-fed" scenarios
(i.e., competitive accretion model, global hierarchical 
model, inertial-inflow model). 
We find that the observed core properties do not match the core properties predicted by the 
turbulent accretion model. On the other hand,  
the competitive accretion, global hierarchical collapse, and inertial-inflow models can partially explain 
 the observed dynamics of dense cores 
(see Section~\ref{sec:msf} for details).  
In general, the observed cores are more consistent 
with the ``clump-fed"  scenario than the ``core-fed" 
scenario. 
However, none of the existing theoretical models can fully 
explain the dynamics of dense cores revealed 
in the ASHES observations.  
A more complete and realistic theoretical model of 
high-mass star formation should reproduce 
the dynamics of dense cores revealed in this work, 
as well as the characteristics of dense cores 
presented in \cite{Sanhueza-2019}.

\begin{acknowledgments}
We thank the anonymous referee for constructive comments that helped to improve this paper.
We thank David Allingham for sharing the NH$_{3}$ rotational excitation temperature maps. We thank Yanett Contreras for her important contribution in shaping this article, contributing with early  analysis of the data and very important insights.  P.S. gratefully acknowledges the support 
from the NAOJ Visiting Fellow Program to have collaborators visiting the National Astronomical Observatory of Japan in November-December 2016. P.S. was financially supported by Grant-in-Aid for Scientific Research (KAKENHI Number  JP22H01271) of Japan Society for the Promotion of Science (JSPS). G.G. acknowledges support by the ANID BASAL project FB210003. 
F.L. acknowledges the support from the National Natural Science Foundation of China grant (12103024),
and the fellowship of China Postdoctoral Science Foundation 2021M691531. 
Data analysis was in part carried out on the open use data analysis computer system at the Astronomy Data Center, ADC, of the National Astronomical Observatory of Japan.  This paper makes use of the following ALMA dataset ADS/JAO.ALMA\#2015.1.01539.S. ALMA is a partnership of ESO (representing its member states), NSF (USA) and NINS (Japan), together with NRC (Canada) and NSC and ASIAA (Taiwan) and KASI (Republic of Korea), in cooperation with the Republic of Chile. The Joint ALMA Observatory is operated by ESO, AUI/NRAO and NAOJ.
\end{acknowledgments}


\vspace{5mm}
\facilities{ALMA}

\software{ CASA \citep{McMullin-2007}, 
Astropy \citep{2013A&A...558A..33A}, 
Matplotlib \citep{4160265}, 
PySpecKit \citep{2011ascl.soft09001G},
SciPy \cite{2020SciPy-NMeth}, 
Seaborn \citep{Waskom2021}, 
LINMIX \citep{Kelly2007}. 
  }


%\software{}



%\appendix

\bibliography{ref}{}
\bibliographystyle{aasjournal}



\appendix

\section{Additional Figures and tables}
\label{app:A}

%%-----------------------------------------------
\begin{figure*}
\centering
\includegraphics[width=0.45\textwidth]{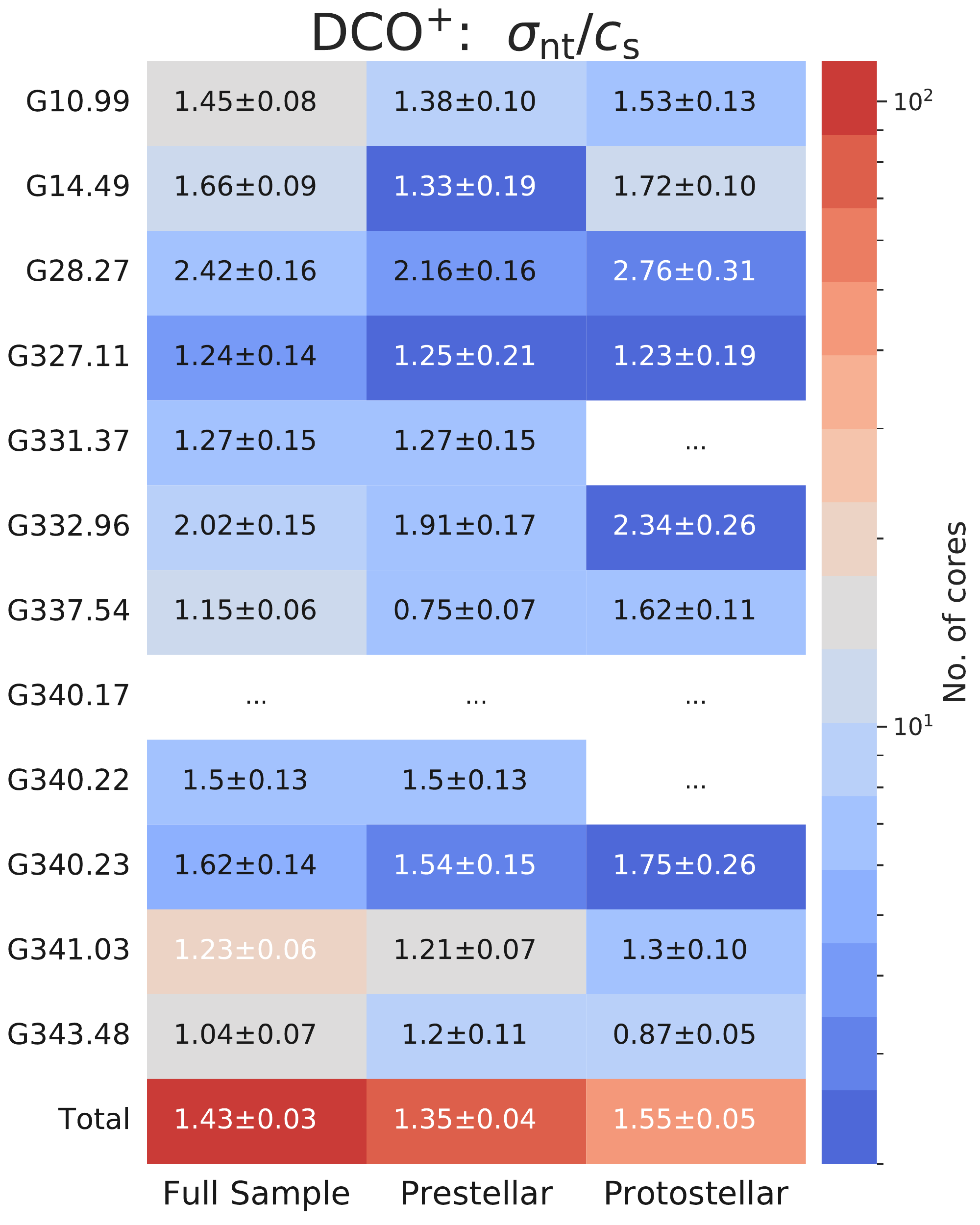}
\includegraphics[width=0.45\textwidth]{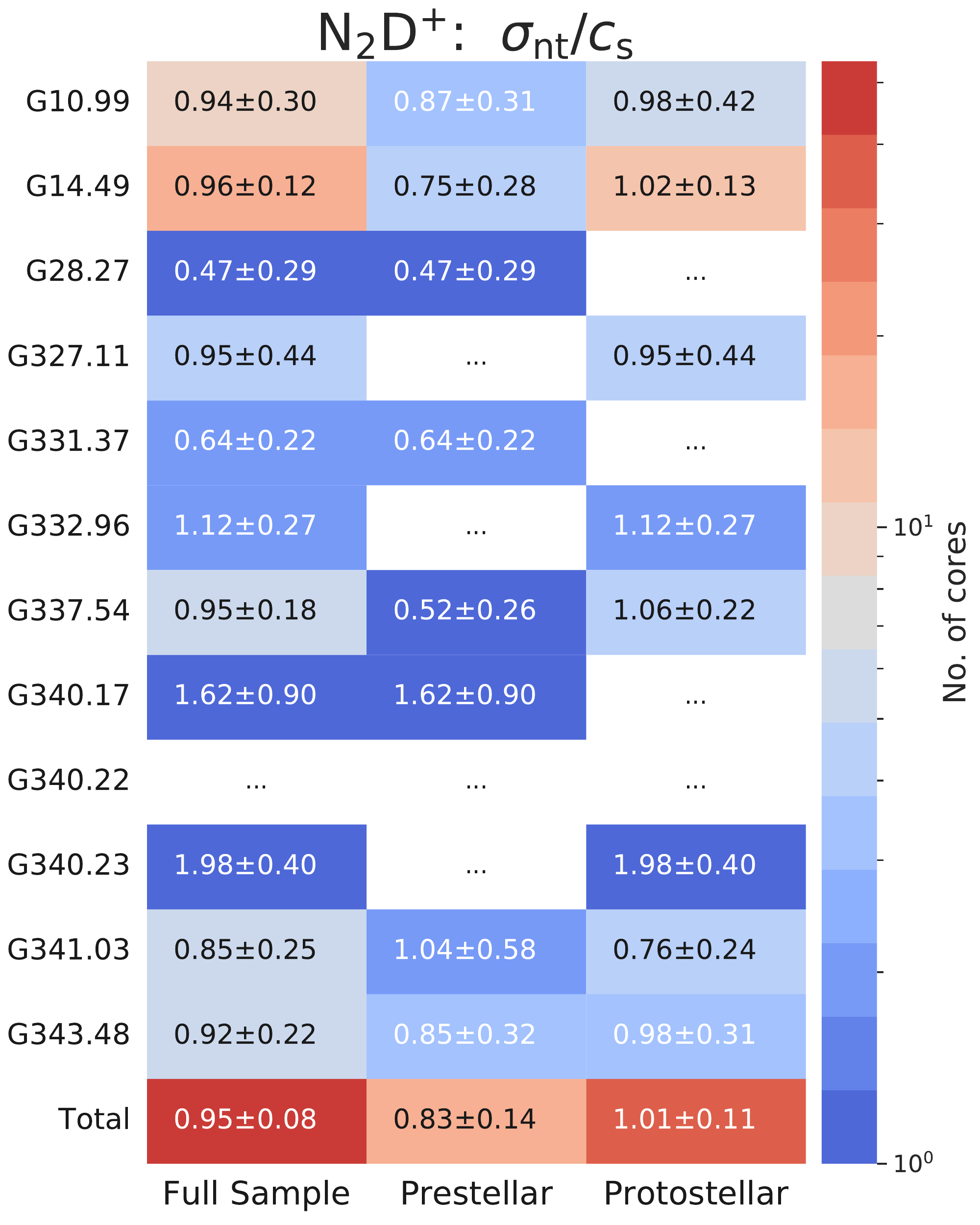}
\caption{Nonthermal velocity dispersion of the cores 
$\sigma_{\rm nt}$, obtained from their \dcop (left) and \n2dp 
(right) molecular line emission for each clump. The total 
row shows the mean value of the velocity dispersion for 
the clumps that have a measurement in both the 
prestellar and protostellar sample. The color scale shows 
the number of cores considered to determine the velocity 
dispersion. 
Dashes denote no available core.} 
\label{fig:hmapsig}%
 \end{figure*}
%%-----------------------------------------------

%%-----------------------------------------------
\begin{figure*}
\centering
\includegraphics[width=0.45\textwidth]{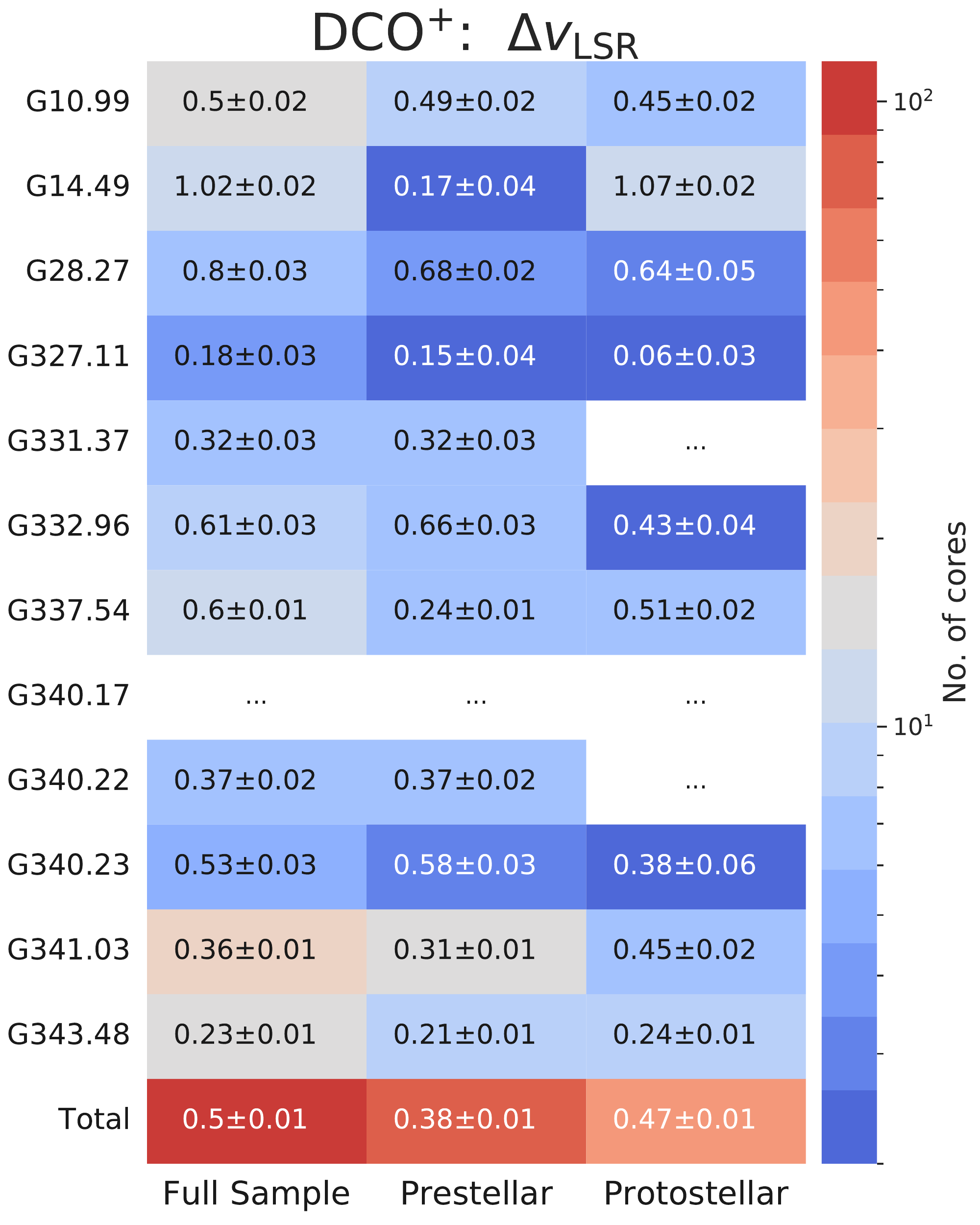}
\includegraphics[width=0.45\textwidth]{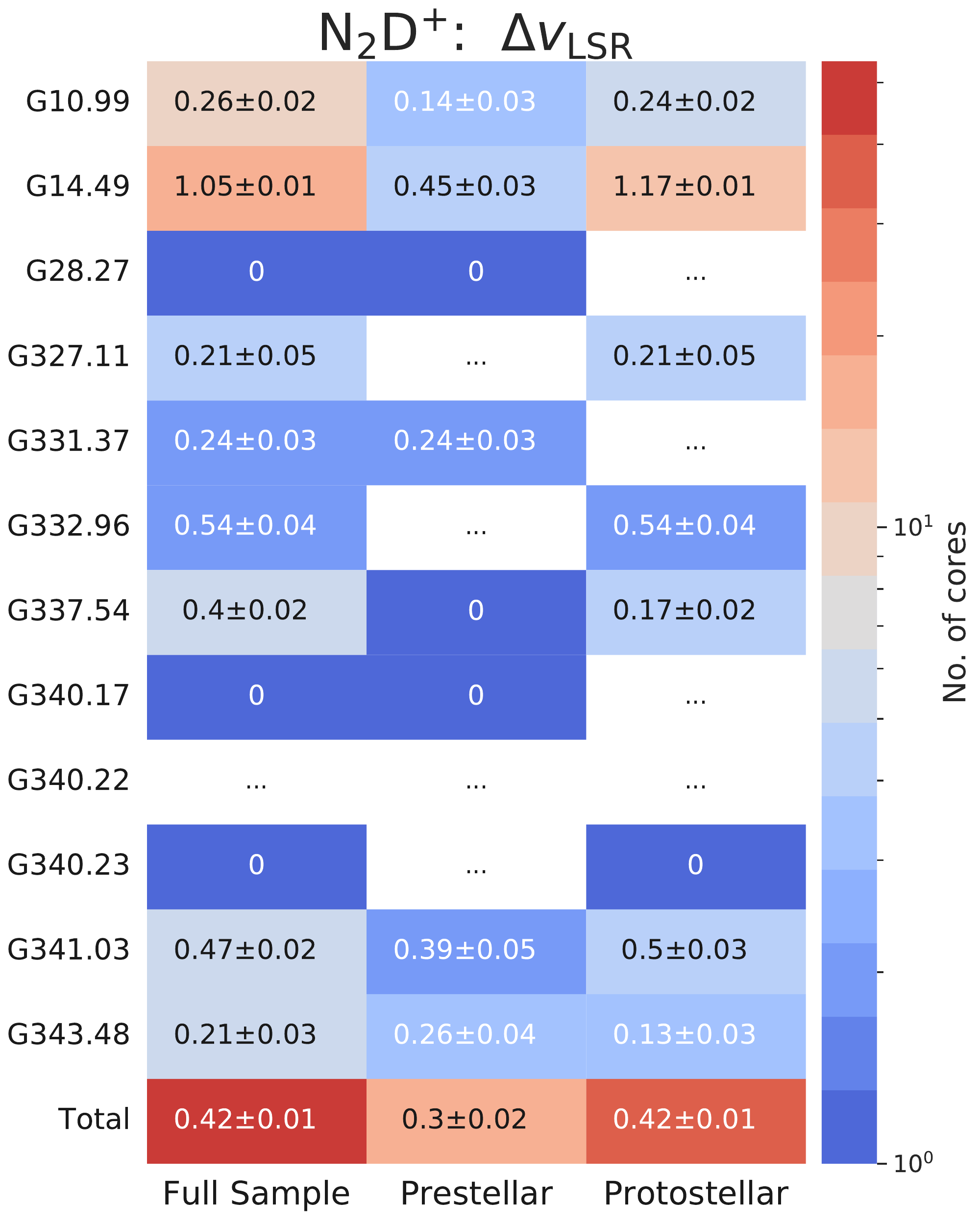}
\caption{Core-to core velocity dispersion of the cores 
$v_{\rm LSR}$ obtained from their \dcop (left) and 
\n2dp (right) molecular line emission for each clump. 
The total row shows the mean value of the core-to-core 
velocity dispersion for the clumps that have a 
measurement in both the prestellar and 
protostellar sample. The color scale shows the number 
of cores considered to determine the core-to-core 
velocity dispersion. `0' means only one available core.
Dashes denote no available core.
} 
\label{fig:hmapvlsr}%
\end{figure*}
%%-----------------------------------------------

%%------------------------------------------------
\begin{figure*}
\centering
\includegraphics[trim={0cm 0cm 0cm 0cm}, 
clip,width=0.85\textwidth]{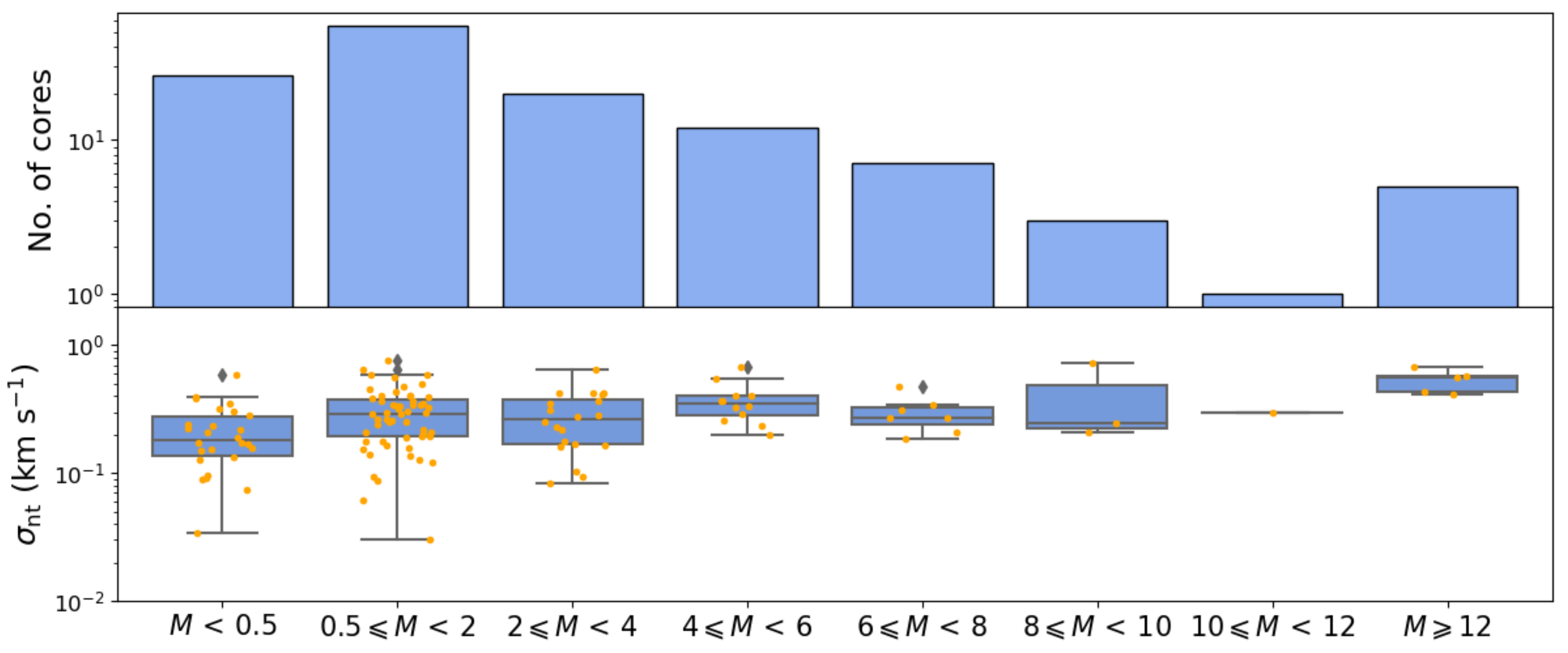}
\caption{Nonthermal velocity dispersion derived 
from the emission of the deuterated species as 
function of the mass of the cores. In the upper 
panel is shown the number of cores per mass bin, 
and in the lower panel is shown a box plot with the 
mean and range of values of the nonthermal 
velocity dispersion per mass bin.
} \label{fig:signth}%
 \end{figure*}
%%------------------------------------------------

%%------------------------------------------------
\begin{figure*}
\centering
\includegraphics[trim={0cm 0cm 0cm 0cm}, 
clip,width=0.6\textwidth]{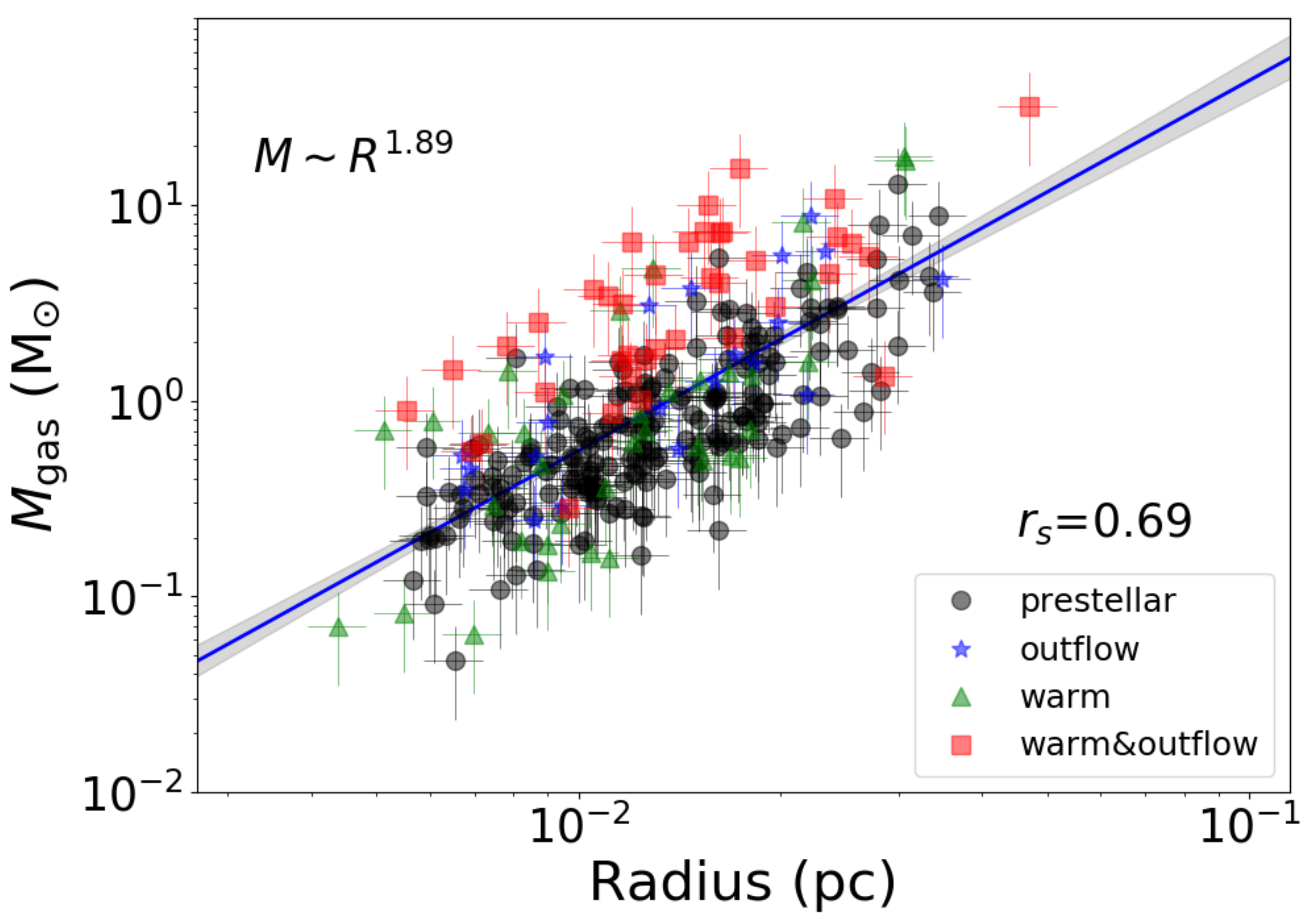}
\caption{The core mass as a function of core radius. 
The blue solid line shows the linear regression to 
all the points in the plot, and the gray shadowed area 
shows the 1$\sigma$ confidence interval for the fit. 
The best fit gives a slope of 1.89 $\pm$ 0.11. 
The Spearman’s rank coefficient is 0.69 with 
the $p$-value of 1.0$\times 10^{-43}$. 
} \label{fig:MR}%
 \end{figure*}
%%------------------------------------------------

\end{document}

%% file: author.tex
%\correspondingauthor{Shanghuo Li}
%\email{shanghuo.li@gmail.com, li@mpia.de}

\author[0000-0003-1275-5251]{Shanghuo Li }
\affiliation{Max Planck Institute for Astronomy, K\"onigstuhl 17, 69117 Heidelberg, Germany}

\author[0000-0002-7125-7685]{Patricio Sanhueza}
\affiliation{National Astronomical Observatory of Japan, National Institutes of Natural Sciences, 2-21-1 Osawa, Mitaka, Tokyo 181-8588, Japan}
\affiliation{Department of Astronomical Science, SOKENDAI (The Graduate University for Advanced Studies), 2-21-1 Osawa, Mitaka, Tokyo 181-8588, Japan}

\author[0000-0003-2384-6589]{Qizhou Zhang}
\affiliation{Center for Astrophysics $|$ Harvard \& Smithsonian, 60 Garden Street, Cambridge, MA 02138, USA}

\author[0000-0003-1649-7958]{Garay, Guido}
\affiliation{Departamento de Astronom\'ia, Universidad de Chile, Las Condes, Santiago, Chile}

\author[0000-0002-6428-9806]{Giovanni Sabatini}
\affiliation{INAF - Istituto di Radioastronomia - Italian node of the ALMA Regional Centre (It-ARC), Via Gobetti 101, I-40129 Bologna, Italy}

\author[0000-0002-6752-6061]{Kaho Morii}
\affiliation{Department of Astronomy, Graduate School of Science, The University of Tokyo, 2-21-1, Osawa, Mitaka, Tokyo 181-0015, Japan}
\affiliation{National Astronomical Observatory of Japan, National Institutes of Natural Sciences, 2-21-1 Osawa, Mitaka, Tokyo 181-8588, Japan}

\author[0000-0003-2619-9305]{Xing Lu}
\affiliation{Shanghai Astronomical Observatory, Chinese Academy of Sciences, 80 Nandan Road, Shanghai 200030, People's Republic of China} 

\author[0000-0002-2149-2660]{Daniel Tafoya}
\affiliation{Department of Space, Earth and Environment, Chalmers University of Technology, Onsala Space Observatory, 439~92 Onsala, Sweden}

\author[0000-0001-5431-2294]{Fumitaka Nakamura}\affiliation{National Astronomical Observatory of Japan, National Institutes of Natural Sciences, 2-21-1 Osawa, Mitaka, Tokyo 181-8588, Japan}
\affiliation{Department of Astronomical Science,  SOKENDAI (The Graduate University for Advanced Studies), 2-21-1 Osawa, Mitaka, Tokyo 181-8588, Japan}

\author[0000-0003-1604-9127]{Natsuko Izumi}
\affiliation{Academia Sinica Institute of Astronomy and Astrophysics, 11F of AS/NTU Astronomy-Mathematics Building, No.1, Section 4, Roosevelt Road, Taipei 10617, Taiwan}

\author[0000-0002-8149-8546]{Ken'ichi Tatematsu}
\affiliation{National Astronomical Observatory of Japan, National Institutes of Natural Sciences, 2-21-1 Osawa, Mitaka, Tokyo 181-8588, Japan}

\author[0000-0002-9832-8295]{Fei Li} 
\affiliation{School of Astronomy and Space Science, Nanjing University, 163 Xianlin Avenue, Nanjing 210023, People's Republic of China}